\documentclass[%
 twocolumn,
 10pt,
superscriptaddress,
 amsmath,amssymb,
 aps,
 pra,
floatfix,
]{revtex4-2}

\usepackage{graphicx}% Include figure files
\usepackage{bm}% bold math
\usepackage{times}
\usepackage{amsmath,amssymb}
\usepackage{enumerate}
\usepackage{multirow}

\usepackage[colorlinks=true,linkcolor=blue,citecolor=red, linktocpage=true,breaklinks=true]{hyperref}

\usepackage{epstopdf}

\newcommand{\eq}{\begin{equation}}
\newcommand{\en}{\end{equation}}
\newcommand{\eqa}{\begin{eqnarray}}
\newcommand{\ena}{\end{eqnarray}}
\newcommand{\tr}{\mathrm{Tr}}

\begin{document}

\title{Asymmetric steerability of quantum equilibrium and nonequilibrium steady states through entanglement detection}% Force line breaks with \\

\author{Kun \surname{Zhang}}
%\email{Email: kun.h.zhang@stonybrook.edu}
\affiliation{Department of Chemistry, State University of New York at Stony Brook, Stony Brook, New York 11794, USA}
\author{Jin \surname{Wang}}
\email{Email: jin.wang.1@stonybrook.edu}
\affiliation{Department of Chemistry, State University of New York at Stony Brook, Stony Brook, New York 11794, USA}
\affiliation{Department of Physics and Astronomy, State University of New York at Stony Brook, Stony Brook, New York 11794, USA}

\date{\today}

\begin{abstract}
	
	Einstein-Podolsky-Rosen steering describes a quantum correlation in addition to entanglement and Bell nonlocality. However, conceptually different from entanglement and Bell nonlocality, quantum steering has an asymmetric definition. Motivated by the asymmetric definition of quantum steering, we study the steerability of two-interacting qubits, which have asymmetric energy levels, coupled with asymmetric environments. The asymmetric (nonequilibrium) environments are two environments with different temperatures or chemical potentials. The Bloch-Redfield equation is applied to study the dynamics of two qubits and its long-time behavior. In our study, the steady-state steerability is determined by an experimentally friendly steering criteria, which demonstrates steering through the entanglement detection. Our results show that the steady states of two asymmetric qubits have the advantage for one direction of steering, compared to the symmetric setup. We also provide analytical results on the minimal coupling strength between the two qubits in order to be steerable. The asymmetric steerability is collectively determined by the nature of the two qubits and the influence from equilibrium or nonequilibrium environments. Nonequilibrium environments with the cost of nonzero entropy production can enhance the steerability in one direction. We also show the strict hierarchy of entanglement, steering and Bell nonlocality of the nonequilibrium steady states, which shows a richer structure of steering than entanglement and Bell nonlocality.
	
\end{abstract}

\maketitle

\section{\label{sec:intro} Introduction}

Einstein-Podolsky-Rosen steering or quantum steering, proposed by Schr\"odinger in 1935 \cite{schrodinger35}, refers to the phenomenon that Alice, sharing quantum correlations with Bob, can steer or manipulate Bob's states by performing measurements on her side. Although the concept of quantum steering was established in the last century, the operational meaning of quantum steering has been clarified only in 2007 \cite{WJD07}. More precisely, quantum steering, from Alice to Bob, means that Bob's conditional states can not have a local hidden state (LHS) description. Such informational definition of quantum steering naturally gives rise to various applications of quantum steering, for example, one-side device-independent quantum key distribution \cite{BCWSW12}, one-side device-independent randomness certification \cite{Law14}, and subchannel discrimination \cite{Piani15}. Recent work also shows that the notion of steering separates the abilities of classical and quantum thermal machines \cite{Beyer19}. More applications about quantum steering can be found in the recent comprehensive review \cite{Uola20}.

Although the definition of quantum steering is unambiguous, the detection of quantum steering is still an on-going research topic. The well-accepted steering criteria include linear steering criteria \cite{CJWR09,CA16}, entropic steering criteria \cite{Schneeloch13,Costa18}, and geometric steering criteria \cite{JHAZW15,YJWG18,Ku18,NNG19}. See \cite{Uola20} for more criteria based on different aspects. Numerous experiments have demonstrated the steerability of two-qubit states, based on different criteria, in the last decade \cite{Saunders2010,Tischler2018,Wollmann20,Yang2020,Bian20,Zhao20}. 

Another two types of well-known quantum correlations are entanglement \cite{Horodecki09} and Bell nonlocality \cite{Brunner14}. These three types of correlations are equivalent for pure states, namely, entangled pure states are both Bell nonlocal and steerable from both sides. However, entanglement, quantum steering and Bell nonlocality have a strict hierarchy for mixed states \cite{Quintino15}. Mixed states are part of pure states in a extended Hilbert space, which suggests that the system is correlated with environments, known as decoherence \cite{Wojciech03}. In general, the environmental effects on correlated systems (through a quantum channel description) have a destructive influence on the steerability \cite{CBA16,SWSY17,PCHLMK19}. 

Interactions between the system and environments are inevitable in real experiments. The question of how to protect quantum correlations from environmental backgrounds is central for designing quantum information processing devices. One idea is to exploit the thermal excitation: entangled excited states can maintain nonzero populations because of interactions between the system and environments \cite{ABV01,KS02}. Moreover, biased environments (two qubits coupled with two different environments having different temperatures or chemical potentials) can further enhance quantum correlations, such as quantum discord \cite{WS11,WW19}, entanglement \cite{QRRP07,LAB07,WWW19}, temporal quantum correlation \cite{CRQ13,ZWW20}, and Bell nonlocality \cite{ZW20}. Although the enhanced entanglement and Bell nonlocality qualitatively suggest that quantum steering should have similar behaviors, the asymmetric definition in quantum steering has no counterpart in the definition of entanglement and Bell nonlocality \cite{Bowles14}. Note that entanglement, quantum steering, and Bell nonlocality are different resources (with different free operations) \cite{SFKSWS20,Gallego15}, which implies that they may have quantitatively different responses to the same environments.  

We study the steerability of two nondegenerate interacting qubits coupled with two individual environments (with the same or different temperatures or chemical potentials) in the steady state regime. We choose the steering criteria based on entanglement detection \cite{Chen18,DSR19}, which is not only experimentally friendly, but also designed to capture the asymmetric features of steering. Note that the well-studied linear steering inequalities fail to characterize the asymmetric steerability of mixed partially entangled states \cite{Chen13}. We apply the Bloch-Redfield equation \cite{Bloch57,Redfield57} to describe the two-qubit evolution, and obtain the nonequilibrium steady states. The Bloch-Redfield  equation shows a more accurate characterization on the coherence of nonequilibrium steady states, compared to the Lindblad form \cite{ZW14,LCS15,Tupkary21,Diniz21}.

%There are two phases of the two interacting qubits \cite{QRRP07,ZW20}: relatively weak interaction (between the two qubits) with separable ground state or relatively strong interaction with entangled ground state. The thermal entanglement (bosonic setups), for the weak interacting qubits, does not support the quantum steering between the two qubits (according to both the LSI up to six measurement choices and the SDED criterion). Strong interacting qubits can demonstrate the quantum steering, if the environments are below the threshold temperatures. Nonequilibrium bosonic environments can promote the quantum steering if the two qubits are nondegenrate (with different frequencies). We also include the fermionic setup: two quantum dots (with the relatively weak interaction) coupled with two leads with the same or different chemical potentials. When the system and the environments are around the resonance (the average qubit frequency equals to the average chemical potential), the population of entangled excited states is maximized, therefore the steerability can be demonstrated. The nonequilibrium fermionic environments also can enhance the steerability if the frequencies of the two qubits are detuned.

The asymmetric steerability is rooted in the asymmetric correlations. For example, Bell diagonal states (classical mixtures of four Bell states) do not show asymmetric steerabilities. However, mixed states composed of partially entangled states can have extreme asymmetric steerabilities (one-way steering) \cite{Bowles16}. In our study, the necessary condition for the asymmetric steerability is having two qubits with different energy levels, where the partially entangled states are eigenstates of the system. We analytically prove that the thermal entanglement of two symmetric qubits do not have the steerability in any direction (based on the entanglement-detection steering criterion). However, the thermal entanglements of two asymmetric qubits can have one direction of steerability. Our study shows that the nonequilibrium environment, with a positive entropy production rate as the thermodynamic cost, can be constructive for one direction of steering. We obtain the hierarchy of entanglement, steering and Bell nonlocality of the nonequilibrium steady states, where the asymmetric features are missing in definitions of entanglement and Bell nonlocality.

The paper is structured as follows. Sec. \ref{sec:steer} reviews the concept of quantum steering as well as the entanglement-detection steering criterion. Sec. \ref{sec:model} introduces the two-qubit model and its dynamical equation. The steerability of equilibrium steady states and nonequilibrium steady states are studied in Secs. \ref{sec:equilibrium} and \ref{sec:nonequilibrium} respectively. The hierarchy structures of quantum correlations of nonequilibrium steady states are shown in Sec. \ref{sec:comparison}. The final section is the conclusion. The appendixes present the explicit forms of the Bloch-Redfield equations as well as their equilibrium steady-state solutions.

\section{\label{sec:steer} Quantum steering and steerability criterion}

\subsection{Quantum steering}

Suppose that Alice and Bob share quantum correlations characterized by the two-qubit density matrix $\rho_{AB}$. Alice's task is to convince Bob that she can steer his state by measuring her qubit. Denote Alice's measurement operator as $E_{a|x}$ with measurement settings $x$ and outcomes $a$. Bob gets the conditional state
\begin{equation}
\rho_{a|x} = \tr_A\left[\left(E_{a|x}\otimes 1\!\!1_2\right)\rho_{AB}\right],
\end{equation}
with the identity operator $1\!\!1_2$. Note that the conditional state is not normalized. The probability for such measurement result is given by $p(a|x) = \tr_{B}(\rho_{a|x})$. The reduced density matrix of Bob is $\rho_{B} = \sum_a\rho_{a|x}$, which is independent of Alice's measurements, required by the no-signaling principle. 

If Bob's conditional state $\rho_{a|x}$ can be explained by some preexisting states with some probability distributions unknown to Bob (the LHS model), Bob's states can be simulated by single-qubit states without any quantum correlations. Specifically, the LHS model says
\begin{equation}
\label{eq LHS form}
    \rho^\text{LHS}_{a|x} = \int d\lambda p(\lambda)p(a|x,\lambda)\varrho_\lambda.
\end{equation}
Here $p(\lambda)$ is the initial probability distribution with the parameter $\lambda$; $p(a|x,\lambda)$ is the probability of measurements result in terms of the parameter $\lambda$; $\varrho_\lambda$ is the preexisting states depending on the distribution of $\lambda$. Alice can cheat Bob by taking the advantage of her knowledge on the hidden variable $\lambda$ and Bob gets only single-qubit states without entanglement. If no such LHS model exists, Bob concludes that Alice can steer his state \cite{Uola20}. The other direction of steering can be similarly defined. There is no guarantee that Bob can steer Alice's state when Alice can steer Bob's state. 

\subsection{\label{subsec:SED} Detecting steerability through entanglement detection}

In general, how to determine the existence of steering, even for generic two-qubit states, is difficult. The challenge is to rule out all possible LHS descriptions given by any measurements. The entanglement-detection steering criterion solves the converse problem: if the LHS model is admitted, the states must follow certain constraints, which give the necessary condition for steerability \cite{Chen18,DSR19}. In another respect, entanglement, steering and Bell nonlocality have the strict hierarchy for mixed states. The implication is: Bell nonlocality can be indirectly detected by the notion of quantum steering \cite{Chen16}; quantum steering can be indirectly detected by the notion of entanglement \cite{Chen18,DSR19}. 

The steerability from Alice to Bob with the density matrix $\rho_{AB}$ can be witnessed if the density matrix $\rho_{{A}\rightarrow{B}}$, defined as
\begin{equation}
    \rho_{{A}\rightarrow{B}} = \frac {\rho_{AB}} {\sqrt 3}  + \frac{3-\sqrt 3}{3}\left(\frac {1\!\!1_2}{2}\otimes \rho_{B}\right),
\end{equation}
is entangled \cite{DSR19}. Here $\rho_{B}$ is Bob's reduced density matrix, namely $\rho_{B} = \tr_{A}(\rho_{AB})$. The corresponding steerability from Bob to Alice can be detected by entanglement detection of the state $\rho_{{B}\rightarrow{A}}$. The new defined state $\rho_{{A}\rightarrow{B}}$ can be viewed as the original density matrix $\rho_{AB}$ after the depolarizing channel (on Alice's qubit) \cite{NC10}. The above criterion can be generalized into any qudit-qubit state (for detecting the steerability of qudit systems to qubit systems) \cite{Chen18,DSR19}. 

The above steering witness requires one to verify the entanglement of a processed state between Alice and Bob. The entanglement verification requires the trust of two parties. However, quantum steering is usually formulated as a one-sided device-independent scenario, namely Alice's or Bob's quantum
state preparation device is not trustful. The trust on Alice's and Bob's apparatus in the entanglement witness can be replaced by the trust of a referee plus two classical channels, known as the quantum-refereed entanglement tests \cite{Buscemi12}. One party trust in the steering witness can be replaced by the trust of a referee plus one classical and one quantum channel \cite{Cavalcanti13}. Although we can replace the entanglement witness in the above steering criterion by the quantum-refereed entanglement tests, the trust on processing of Alice's qubit is not guaranteed. Nevertheless, we stick on such steering witness requiring the full knowledge of the bipartite state for the theoretical study. It would be interesting to explore the influence of environments on the quantum-refereed steering tests in the future.

Suppose that the two-qubit state $\rho_{AB}$, in the local basis, has the structure
\begin{equation}
\label{def:X_state}
    \rho_{AB} = \left(\begin{array}{cccc}
   \rho_{11}	& 0 & 0 & \rho_{14} \\
0	& \rho_{22} & \rho_{23} & 0 \\
0	& \rho_{23}^* & \rho_{33} & 0 \\
\rho_{14}^*	& 0 & 0 & \rho_{44}
\end{array} \right),
\end{equation}
called the ``X''-state \cite{Yu07}. Then the constructed density matrices $\rho_{{A}\rightarrow{B}}$ and $\rho_{{B}\rightarrow{A}}$ preserve the ``X'' structure. Entanglement criterion, such as the positive partial transpose criterion \cite{Horodecki01}, has a simple closed form for the ``X''-state. If we assume $|\rho_{23}|\gg|\rho_{14}|$ (which is the case in our two-qubit model), then the steerability from Alice to Bob and Bob to Alice satisfy inequalities
\begin{subequations}
\begin{align}
    \label{eq:A_to_B}&|\rho_{23}|^2 > f_a+f_b;\\
    \label{eq:B_to_A}&|\rho_{23}|^2 > f_a-f_b,
\end{align}
\end{subequations}
respectively. Here $f_a$ and $f_b$ are given by population terms of the density matrix:
\begin{subequations}
\begin{multline}
\label{eq:f_a}f_a = \frac{2+\sqrt 3}{2}\rho_{11}\rho_{44}+\frac{2-\sqrt 3}{2}\rho_{22}\rho_{33} \\
    + \frac 1 4 (\rho_{11}+\rho_{44})(\rho_{22}+\rho_{33});
\end{multline}
\begin{equation}
\label{eq:f_b}f_b = \frac 1 4 (\rho_{11}-\rho_{44})(\rho_{22}-\rho_{33}).
\end{equation}
\end{subequations}
Then if $\rho_{11} = \rho_{44}$ or $\rho_{22}=\rho_{33}$, we have the symmetric steerability. If $f_b>0$, then we have the asymmetric steerability from Bob to Alice. Similar inequalities can be derived if $|\rho_{14}|\gg|\rho_{23}|$. 

Entanglement properties of any two-qubit state can be mapped to a counterpart-``X'' state through a unitary transformation, which preserves states' entanglement properties \cite{Mendoncca14}. It is interesting to prove that the steerability of any two-qubit state, based on the entanglement-detection steering criterion, can be witnessed on a corresponding ``X'' state, which is beyond the scope of our current paper. 

The entanglement-detection steering criterion can distinguish the asymmetric steerability based on tests on some types of two-qubit states \cite{DSR19,Yang2020}. Recent experiments show that it is also superior than some other steering criteria which can also reveal the asymmetric steerability \cite{Yang2020}. We also find that the steerability of the system in this paper revealed by the entanglement-detection steering criterion, covers the linear steering criteria up to six measurements \cite{Saunders2010}, where the latter also fails to capture the asymmetric steerability. Therefore, we concentrate on the asymmetric steerability revealed by the entanglement-detection steering criterion in this study. 

\quad

\section{\label{sec:model} Model and Quantum Master Equation}

\subsection{\label{subsec:model} The two-qubit model}

\begin{figure}
    \begin{center}
    	\includegraphics[width=\columnwidth]{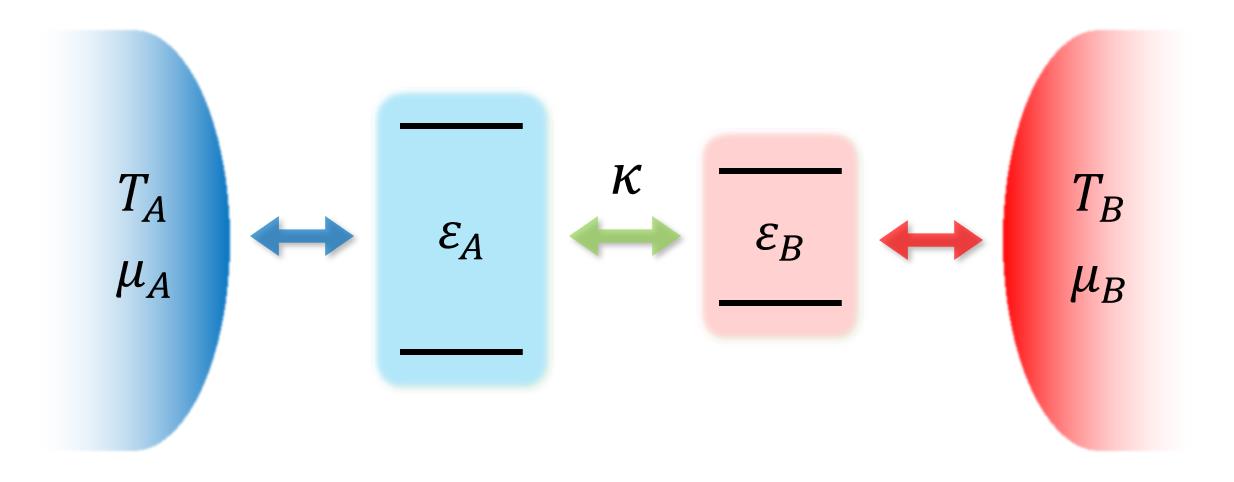}
    	\caption{The interacting two qubits, with different frequencies $\varepsilon_{A}\neq\varepsilon_{B}$, are coupled with two environments which may have different temperatures $T_{A}\neq T_{B}$ or chemical potentials $\mu_{A}\neq \mu_{B}$.}\label{fig model}
    \end{center}
\end{figure}

Consider two qubits, labeled with $A$ and $B$, having the XY interaction \cite{IABDLSS99}, which gives the Hamiltonian
\begin{equation}
\label{def H S}
    \mathcal H_{AB}=\frac{\varepsilon_{A}}{2}\sigma^z_{A}+\frac{\varepsilon_{B}}{2}\sigma^z_{B}+\frac \kappa 4\left(\sigma_{A}^x\sigma_{B}^x+\sigma_{A}^y\sigma_{B}^y\right),
\end{equation}
with the Pauli matrices $\sigma^{x,y,z}$. See Fig. \ref{fig model}. Without loss of generality, we assume $\varepsilon_{A}\geq\varepsilon_{B}$. The XY interaction also describes a dipole-dipole interaction between two atoms with distance $r$, where $\kappa\propto r^{-3}$ \cite{PK02,Wang21}. The spin Hamiltonian $\mathcal H_{S}$ also equivalently describes two-electron sites with the hopping rate $\kappa$ \cite{WW19}. We set the unit $\hbar = 1$. 

The two-qubit system has the local basis $\{|0\rangle_{A(B)},|1\rangle_{A(B)}\}$, which are the eigenbasis of Pauli matrix $\sigma^z_{A(B)}$. The two qubits have eigenstates $|00\rangle_{AB}$ and $|11\rangle_{AB}$ which are product states (no correlations). The other two eigenstates are entangled, given by
\begin{subequations}
\begin{align}
    \label{eq:psi-}&|\psi^-(\theta)\rangle_{AB} = \cos\frac{\theta}{2}|01\rangle_{AB}-\sin\frac{\theta}{2}|10\rangle_{AB};\\
    \label{eq:psi+}&|\psi^+(\theta)\rangle_{AB} = \sin\frac{\theta}{2}|01\rangle_{AB}+\cos\frac{\theta}{2}|10\rangle_{AB}.
\end{align}
\end{subequations}
The angle $\theta\in[0,\pi/2]$, called the detuning angle, is given by $\theta = \arctan[\kappa/(\varepsilon_{A}-\varepsilon_{B})]$. Note that $\theta=\pi/2$, which means the two qubits are symmetric $\varepsilon_{A} = \varepsilon_{B}$, then states $|\psi^-(\pi/2)\rangle_{AB}$ and $|\psi^+(\pi/2)\rangle_{AB}$ are maximally entangled.

There are two quantum phases of the system \cite{QRRP07,ZW20}. When $\kappa<\sqrt{\varepsilon_{A}\varepsilon_{B}}$ (weak-coupling phase), the ground and first excited states are $|00\rangle_{AB}$ and $|\psi^-(\theta)\rangle_{AB}$ respectively. When $\kappa>\sqrt{\varepsilon_{A}\varepsilon_{B}}$ (strong-coupling phase), the ground state becomes the entangled state $|\psi^-(\theta)\rangle_{AB}$. The ground states of these two phases have qualitatively different correlations. We also expect different behaviors of the two phases when the system couples to environments. 

In terms of the two reservoirs, we assume the general free bosonic Hamiltonian given by
\begin{equation}
    \mathcal H_{{R}_{j}} = \sum_{k_{j}} \varepsilon_{k_{j}}a^\dag_{k_{j}}a_{k_{j}},
\end{equation}
where the two reservoirs are labeled with $j=A,B$. The operator $a_k$ follows the commutative relation. When considering the fermionic reservoirs, in which the two qubits represent for two electron sites, we assume the free electronic Hamiltonian, where the operator $a_k$ follows the anticommutative relation.

We consider the spin-boson type interaction \cite{Leggett87} between the system and reservoir, given by
\begin{equation}
\label{def:H_I}
    \mathcal H_{I_j} = \sigma_{j}^x \sum_{k_{j}} \lambda_{k_{j}}(a_{k_{j}}+a^\dag_{k_{j}}).
\end{equation}
Qubit $A(B)$ couples to the reservoir $A(B)$ accordingly. Here $\lambda$ is the coupling strength between the system and reservoir.

%\begin{equation}
%\label{def theta}
%    \theta = \arctan\left(\frac{\kappa}{\varepsilon_{A}-\varepsilon_{B}}\right).
%\end{equation}

\subsection{\label{subsec:BR eq} Dynamics of the system}

The Bloch-Redfield equation is derived based on the weak-coupling approximation and the Born-Markov approximation \cite{BP02}. Note that the weak- or strong- coupling phase described before is only specified for the interaction between two qubits. We assume the weak-coupling approximation (between the system and environment) to be valid both in the weak- and strong- coupling phases. 

For our two-qubit model, the total Hamiltonian is 
\begin{equation}
    \mathcal H = \mathcal H_{AB} + \mathcal H_{R_A}+ \mathcal H_{R_B} + \mathcal H_{I_A} + \mathcal H_{I_B}.
\end{equation}
The Bloch-Redfield equation (in the interaction picture) is obtained by tracing out the two environments denoted as $R_A$ and $R_B$, given by
\begin{widetext}
\begin{equation}
\label{eq:master_eq}
    \frac{\text{d}\rho_{AB}(t)}{\text{d}t}=-\int_0^\infty d\tau \tr_{{R}_{A}{R}_{B}} \left[\mathcal H_{{I}_{A}}(t)+\mathcal H_{{I}_{B}}(t),\left[\mathcal H_{{I}_{A}}(t-\tau)+\mathcal H_{{I}_{B}}(t-\tau),\rho_{S}(t)\otimes\rho_{{R}_{A}}\otimes\rho_{{R}_{B}}\right]\right],
\end{equation}
\end{widetext}
where $\rho_{A(B)}$ is the equilibrium state of environment $A(B)$ with temperature $T_{A(B)}$ and chemical potential $\mu_{A(B)}$.

In the Schr\"odinger's picture, the Bloch Redfield equation has the form
\begin{equation}
    \label{eq:master_eq2}
    \frac{\text{d}\rho_{AB}}{\text{d}t}=i\left[\rho_{AB},H_{AB}\right]+\sum_{j}\mathcal D_j[\rho_{AB}].
\end{equation}
The first term describes the coherent evolution of the system; the second term describes the dissipator caused by the reservoir A or B. To solve the steady state, we can rewrite the master equation Eq. (\ref{eq:master_eq}) into the matrix form, given by
\begin{equation}
\label{def:M}
    \frac{\text{d}}{\text{d}t}|\rho_{S}\rangle = \mathcal M |\rho_{S}\rangle.
\end{equation}
Therefore, the steady state solution is the eigenvector of $\mathcal M$ with eigenvalue 0. The system does not have the dark states in terms of the dynamics \cite{Daniel19}, therefore the unique steady state solution is guaranteed. The elements of the evolution matrix $\mathcal M$ are given in Appendix \ref{App:A}.

In general, it is hard to explicitly characterize the system-environment coupling $\lambda_k$. Instead, knowing the spectral function (also called coupling spectrum), defined by 
\begin{equation}
\label{def gamma}
\gamma_j(\varepsilon)=\pi\sum_{k_j} |\lambda_{k_j}|^2\delta(\varepsilon-\varepsilon_{k_j}),
\end{equation}
is suffice to determine the dynamics of system \cite{Leggett87}. The spectral function can be categorized into different types, such as Ohmic, sub-Ohmic and super-Ohmic. However, research shows that the nonequilibrium steady states given by different types of spectral functions have similar behaviors \cite{WWW19}. Since only steady states are considered in our study, we assume the symmetric constant spectral functions for simplicity, namely $\gamma_{A} = \gamma_{B} = \gamma$. 

The difference between the Lindblad equation and the Bloch-Redfield equation is the secular approximation, which is applied in the Lindblad \cite{BP02}. The secular approximation averages out the cross transitions between different energy levels \cite{LCS15}. However, such transitions not only contribute to oscillations in the dynamics, but also, in the nonequilibrium environments, gives nonzero energy-basis steady-state coherence in the order $\mathcal O(\gamma^2)$ \cite{WW19,WWW19}. Recent study also shows that such coherence is necessary for the local conservation laws in the nonequilibrium environments \cite{Tupkary21}. The issue of Bloch-Redfield equation is sometimes lack of positivity, especially in the fermionic setup \cite{WWW19}. We avoid the issue by choosing the range of parameters which give the positive defined steady states.

\section{\label{sec:equilibrium} Steerability of equilibrium steady states}

\subsection{Bosonic equilibrium environments}

\begin{figure}
    \begin{center}
    	\includegraphics[width=\columnwidth]{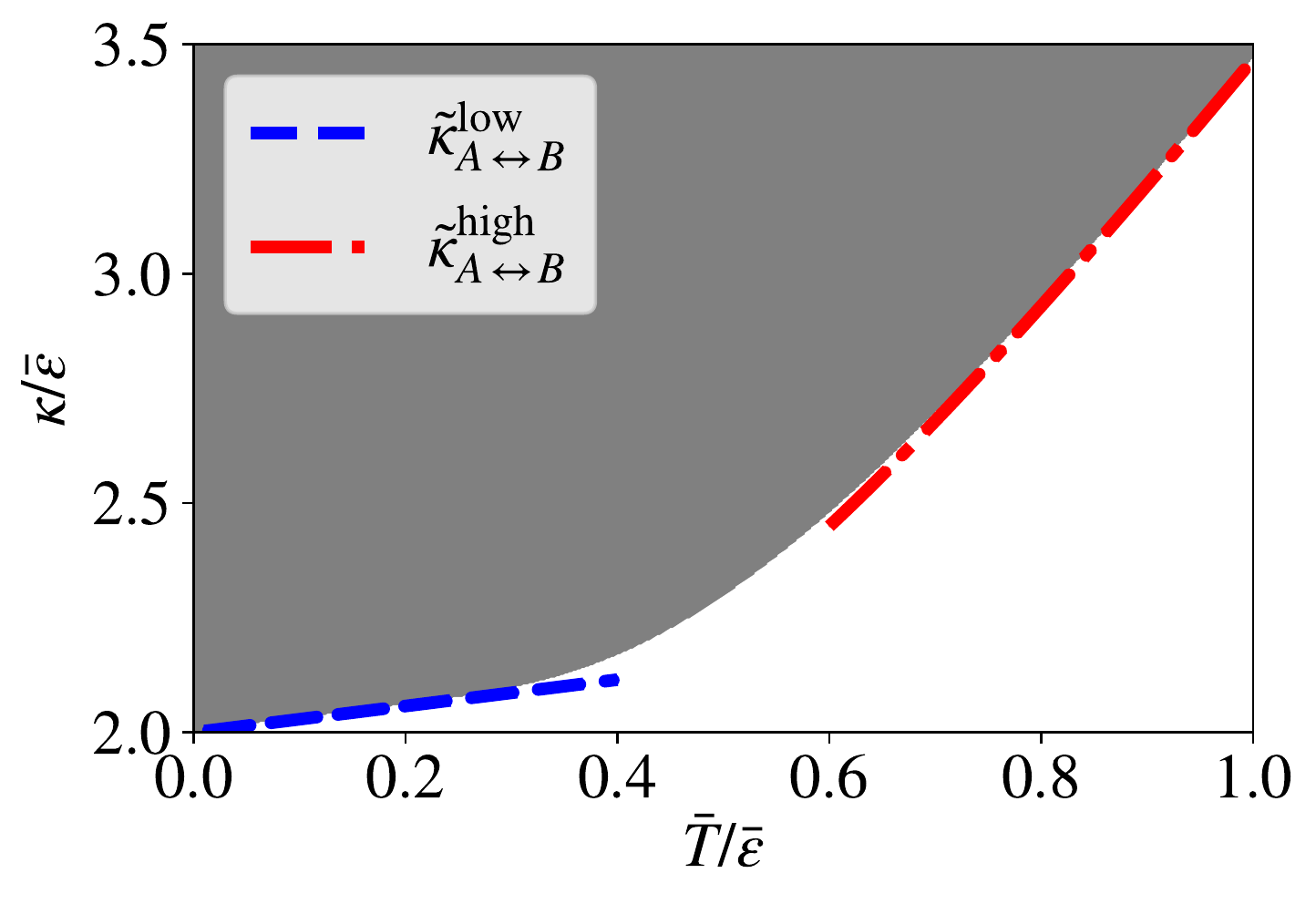}
    	\caption{Steerable regions (grey color) in terms of inter-qubit coupling strength $\kappa$ and equilibrium temperature $\bar T$. The threshold coupling strengths $\tilde\kappa^\text{low}_{A\leftrightarrow B}$ and $\tilde\kappa^\text{high}_{A\leftrightarrow B}$ are given by Eqs. (\ref{eq:kappa_low}) and (\ref{eq:kappa_high}) respectively.}\label{fig k_T_eq}
    \end{center}
\end{figure}

The steady state under equilibrium environments ($T_A = T_B = \bar T$) can either be obtained by solving the master equation, or imposing the eigenstate thermalization. The thermal reduced density matrix has the form $\rho_{AB}^\text{ss} = e^{-\beta\mathcal H_{AB}}/\mathcal Z$ with the partition function $\mathcal Z$ and the inverse temperature $\beta = 1/T$. We set Boltzmann constant as 1. The thermal density matrix is also the steady state solution given by the Bloch-Redfield equation in Eq. (\ref{def:M}). See Appendix \ref{App:B} for the steady state solutions, either in the energy basis or local basis. 

If the two qubits are symmetric $\varepsilon_{A} = \varepsilon_{B}=\bar\varepsilon$, we expect symmetric steerability between $A$ and $B$. More specifically, the population of local states $|01\rangle_{AB}$ and $|10\rangle_{AB}$ are the same, which corresponds to $f_b = 0$ defined in Eq. (\ref{eq:f_b}). Therefore, the two steerability inequalities presented in Eqs. (\ref{eq:A_to_B}) and (\ref{eq:B_to_A}) are identical. Both the weak- and strong- coupling phases give the steerability inequality
\begin{equation}
\label{eq:sy_inequality}
    \frac{\sqrt 3}{2}\cosh^2\left(\frac{\beta\kappa}{2}\right)-\cosh(\beta\bar\varepsilon)\cosh\left(\frac{\beta\kappa}{2}\right)>\frac{4+\sqrt 3}{2},
\end{equation}
derived from Eqs. (\ref{eq:A_to_B}) and (\ref{eq:B_to_A}). It is easy to see that there is no solution in the weak-coupling phase ($\kappa<2\bar\varepsilon$).

The symmetric steerability exists in the strong-coupling phase. We numerically find the steerable regions in terms of the parameters $\kappa$ and $T$ in Fig. \ref{fig k_T_eq}. It is easy to understand that higher temperatures of environments, the two qubits require stronger couplings to maintain the steerability. When the temperature is low ($\beta\bar\varepsilon\gg 0$), the inequality presented in Eq. (\ref{eq:sy_inequality}) gives
\begin{equation}
\label{eq:kappa_low}
    \kappa>\tilde \kappa^\text{low}_{A\leftrightarrow B} = 2\bar\varepsilon+\ln\left(\frac 4 3\right)T.
\end{equation}
Here we define the threshold coupling strength $\tilde \kappa^\text{low}_{A\leftrightarrow B}$, which linearly increases with the temperature $T$. The changing rate is $\ln(4/3)\approx 0.125$. If the temperature is high ($\beta\bar\varepsilon\approx 0$), the steerability inequality gives
\begin{equation}
\label{eq:kappa_high}
    \kappa>\tilde\kappa^\text{high}_{A\leftrightarrow B} = 3.121 T+0.347 \frac{\bar\varepsilon^2}{T}.
\end{equation}
The linearly changing rate becomes $3.121$ now. See Fig. \ref{fig k_T_eq} for comparison between the numerical and analytical results. 

\begin{figure}
    \begin{center}
    	\includegraphics[width=\columnwidth]{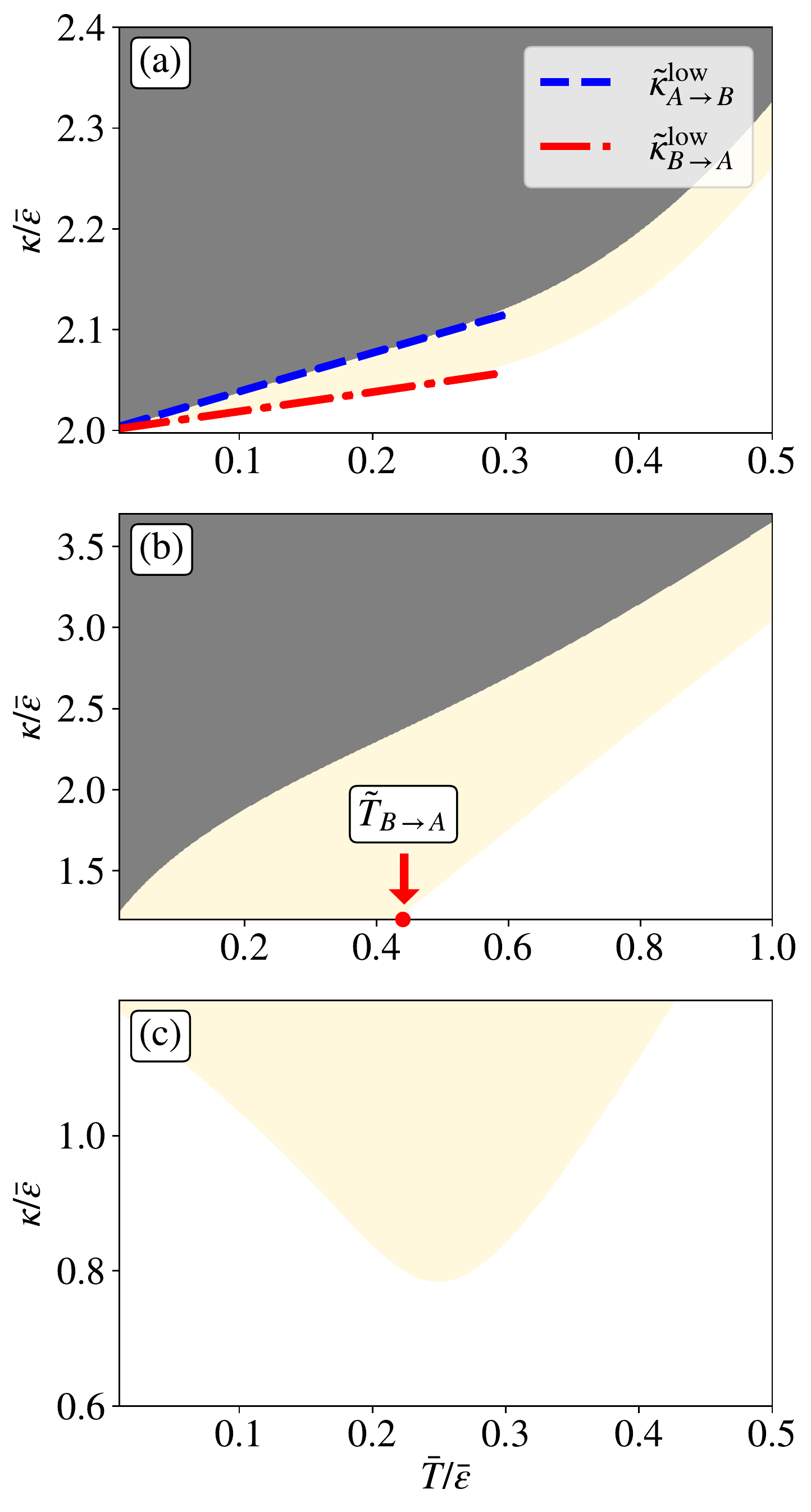}
    	\caption{Parameter regions of two-way steerable (grey) and steerable from Bob to Alice (yellow). The threshold coupling strengths $\tilde \kappa^\text{low}_{A\rightarrow B}$ and $\tilde \kappa^\text{low}_{B\rightarrow A}$ are given by Eqs. (\ref{eq:kappa_A_B}) and (\ref{eq:kappa_B_A}). The range of $\kappa$ in (a)(b) is in the strong-coupling phase; (c) is in the weak-coupling phase. The detuning is set as (a) $\Delta\varepsilon=0.1\bar\varepsilon$; (b)(c) $\Delta\varepsilon=1.6\bar\varepsilon$.}\label{fig k_T_asy_eq}
    \end{center}
\end{figure}

Suppose that the average frequency of two qubits is $\bar\varepsilon = (\varepsilon_A+\varepsilon_B)/2$. If the two qubits are detuned, namely $\Delta\varepsilon=\varepsilon_A-\varepsilon_B>0$, the ground singlet state becomes a partially entangled state. However, the detuned two qubits have larger gap between the ground and first excited states. The gap is given by $\sqrt{\Delta\varepsilon^2+\kappa^2}-\bar\varepsilon $. Therefore the two qubits become more robust to the environment, correspondingly entanglement or Bell nonlocality maybe stronger \cite{ZW20}. The detuned qubits are no longer symmetric in terms of $A$ and $B$. The environment affects differently on these two directions of steering. At low temperature ($\beta\bar\varepsilon\gg 0$) and small detuning $\Delta\varepsilon/\bar\varepsilon\ll 1$ regimes, we have the different threshold coupling strengths for different directions of steering:
\begin{subequations}
\begin{align}
\label{eq:kappa_A_B}
    \tilde \kappa^\text{low}_{A\rightarrow B} = & \tilde \kappa^\text{low}_{A\leftrightarrow B} + \frac{2\Delta\varepsilon}{\tilde \kappa^\text{low}_{A\leftrightarrow B}}T;\\
\label{eq:kappa_B_A}
    \tilde \kappa^\text{low}_{B\rightarrow A} = & \tilde \kappa^\text{low}_{A\leftrightarrow B} - \frac{2\Delta\varepsilon}{\tilde \kappa^\text{low}_{A\leftrightarrow B}}T,
\end{align}
\end{subequations}
where $\tilde\kappa^\text{low}_{A\leftrightarrow B}$ is the threshold coupling strength at the symmetric case ($\Delta\varepsilon = 0$), defined in Eq. (\ref{eq:kappa_low}). Therefore steering from $B$ to $A$ is easier than the other direction. See Fig. \ref{fig k_T_asy_eq} for the numerical results. 

Previous studies have constructed the one-way steerable state by adding local noises on partially entangled states \cite{Bowles16,Yang2020}. In our two-qubit model, suppose that the four eigenstates $|00\rangle_{AB}$, $|11\rangle_{AB}$ and $|\psi^\pm(\theta)\rangle_{AB}$ have the probabilities $p_{00}$, $p_{11}$ and $p_\pm$. The imbalance factor $f_b$, defined in the steerability inequalities in Eqs. (\ref{eq:A_to_B}) and (\ref{eq:B_to_A}), has the analytical form
\begin{equation}
\label{eq:f_b_detail}
    f_b = (p_{00}-p_{11})(p_--p_+)\cos\theta.
\end{equation}
In the strong-coupling phase, the thermal density matrix always gives $p_->p_{00}>p_{11}>p_+$. Then the sign of $f_b$ is completely determined by the detuning angle $\theta$. When $\varepsilon_A>\varepsilon_B$, we have $\theta<\pi/2$, which gives a positive $f_b$. Therefore the steerability from Bob to Alice is favored. Here we can also intuitively understand that Bob's local state subjects to more noises from the environment $B$ if $\varepsilon_B<\varepsilon_A$, then the steerability from Alice to Bob becomes harder. 

The general trend for increasing $\Delta\varepsilon$ is to increase $\tilde\kappa_{A\rightarrow B}$ while decrease $\tilde\kappa_{B\rightarrow A}$, see Fig. \ref{fig k_T_asy_eq}. More interestingly, if the detuning is larger than a threshold $\Delta\tilde\varepsilon_{B\rightarrow A}$, namely
\begin{equation}
\label{eq:delta_w}
    \Delta\varepsilon > \Delta\tilde\varepsilon_{B\rightarrow A} = \frac{4\sqrt 3-6}{3}\bar\varepsilon,
\end{equation}
the threshold coupling strength $\tilde\kappa_{B\rightarrow A}$ is smaller than $2\sqrt{\varepsilon_A\varepsilon_B}$. This suggests that there is a robust temperature $\tilde T_{B\rightarrow A}$, giving the steerability from Bob to Alice below $\tilde T_{B\rightarrow A}$ in the strong-coupling phase. Besides, this also suggests that steering from Bob to Alice is possible in the weak-coupling phase if $\Delta\varepsilon > \Delta\tilde\varepsilon_{B\rightarrow A}$. See Fig. \ref{fig k_T_asy_eq} for the numerical verification.

\subsection{Fermionic equilibrium environments}

\begin{figure}
    \begin{center}
    	\includegraphics[width=\columnwidth]{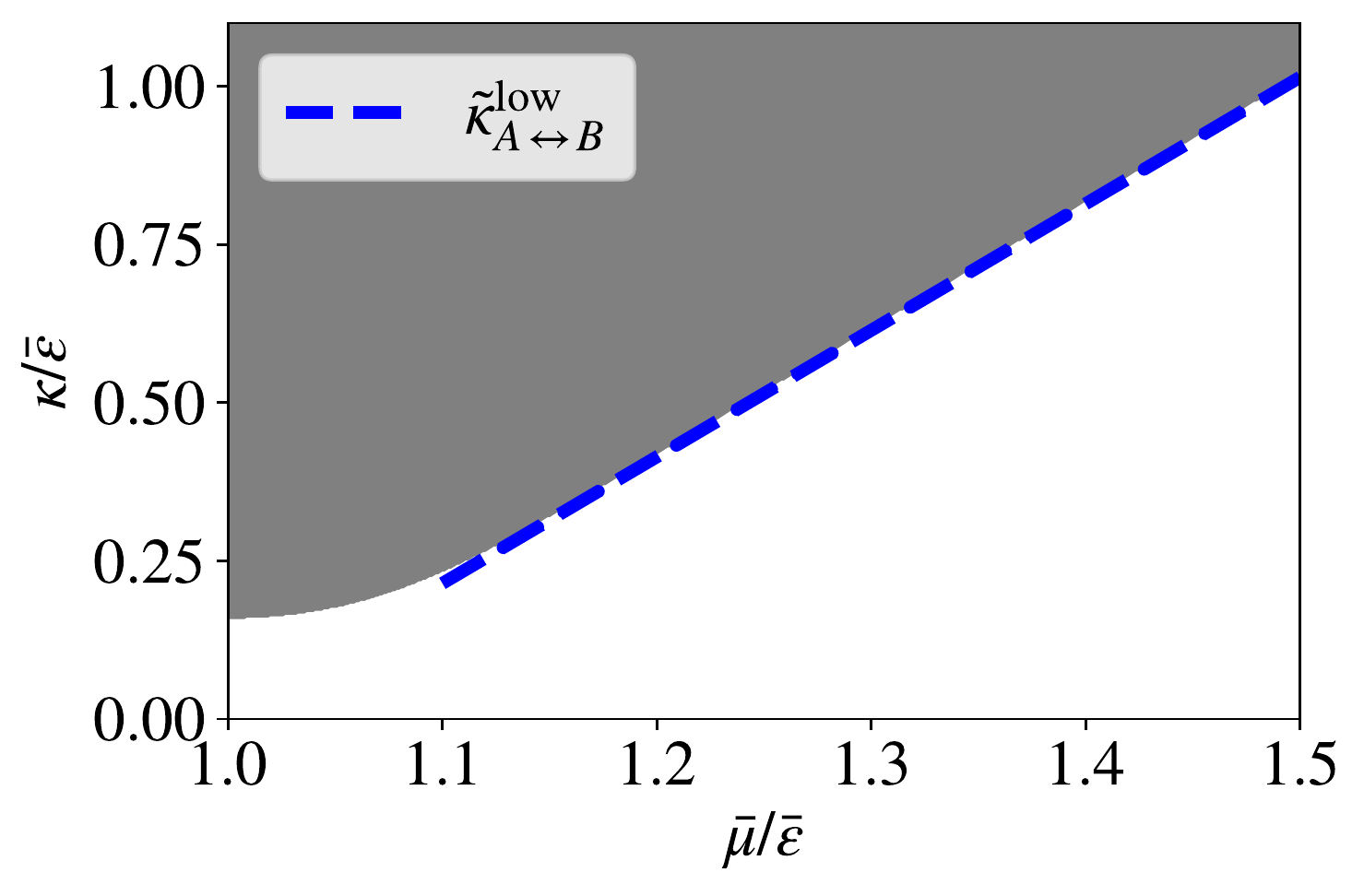}
    	\caption{Steerable regions (grey color) in terms of coupling strength $\kappa$ and equilibrium chemical potential $\bar\mu$. The threshold coupling $\tilde\kappa_{A\leftrightarrow B}$ is shown in Eq. (\ref{eq:kappa_mu_w}). Temperature is set as $T = 0.05\bar\varepsilon$.}\label{fig k_mu_eq}
    \end{center}
\end{figure}

We consider the fermionic system described by the Hamiltonian $\mathcal H_S$ in Eq. (\ref{def H S}). Note that it is a toy fermionic model, neglecting the Coulomb interaction between two electrons. Nevertheless, the model can qualitatively predict the difference between bosonic and fermionic environmental influences on the system.

In the fermionic setup, the equilibrium steady state $\mu_A=\mu_B=\bar\mu$ is predicted by the grand canonical ensemble, given by $\rho_{AB}^\text{ss} = e^{-\beta(\mathcal H_{AB}-\bar\mu\mathcal N_{AB})}$ with the number operator $\mathcal N_{AB}$. It is also the steady-state given by the Bloch-Redfield equation in Eq. (\ref{def:M}). See Appendix \ref{App:B} for the explicit expression of the steady state solution in the energy and local bases. We only consider the weak-coupling phase in the fermionic setup, since the degree of tunneling is relatively weak \cite{Oosterkamp98}.

For the symmetric two qubits $\varepsilon_A=\varepsilon_B$, the two steerability inequalities give
\begin{equation}
\label{eq:f_sy_inequality}
    \frac{\sqrt 3}{2}\cosh^2\left(\frac{\beta\kappa}{2}\right)-\cosh\left(\beta(\bar\varepsilon-\bar\mu)\right)\cosh\left(\frac{\beta\kappa}{2}\right)>\frac{4+\sqrt 3}{2}.
\end{equation}
As hyperbolic cosine function is even, chemical potentials $\mu$ and $2\bar\omega-\mu$ give the same steerability of the system. The above inequality has the solution in the weak-coupling phase ($\kappa<2\bar\varepsilon$). For example, at the resonant point $\bar\mu = \bar\varepsilon$, we have the threshold coupling strength
\begin{equation}
\label{eq:kappa_mu_w}
    \kappa > \tilde\kappa_{A\leftrightarrow B}(\mu = \bar\varepsilon) = 2\text{Arccosh}\left(\frac{\sqrt 3 + 2\sqrt{3+3\sqrt 3}}{3}\right)T,
\end{equation}
with the inverse hyperbolic cosine function $\text{Arccosh}$. As $\beta|\bar\varepsilon-\bar\mu|\gg 0$ (low temperature case), we find the threshold coupling strength given by
\begin{equation}
\label{eq:kappa_f_away}
    \kappa > \tilde\kappa^\text{low}_{A\leftrightarrow B} = 2|\bar\varepsilon-\bar\mu|+\ln\left(\frac 4 3\right)T.
\end{equation}
Therefore, away from the resonant point $\bar\mu = \bar\varepsilon$, the threshold coupling strength linearly increases with the chemical potential. The resonant point $\bar\mu = \bar\varepsilon$ maximizes the populations of the first and second excited states, which are maximally entangled. This is consistent with the behavior of steady state entanglement and Bell nonlocality \cite{WWW19,ZW20}. The regions of symmetric steerability are shown in Fig. \ref{fig k_mu_eq}.

\begin{figure}
    \begin{center}
    	\includegraphics[width=\columnwidth]{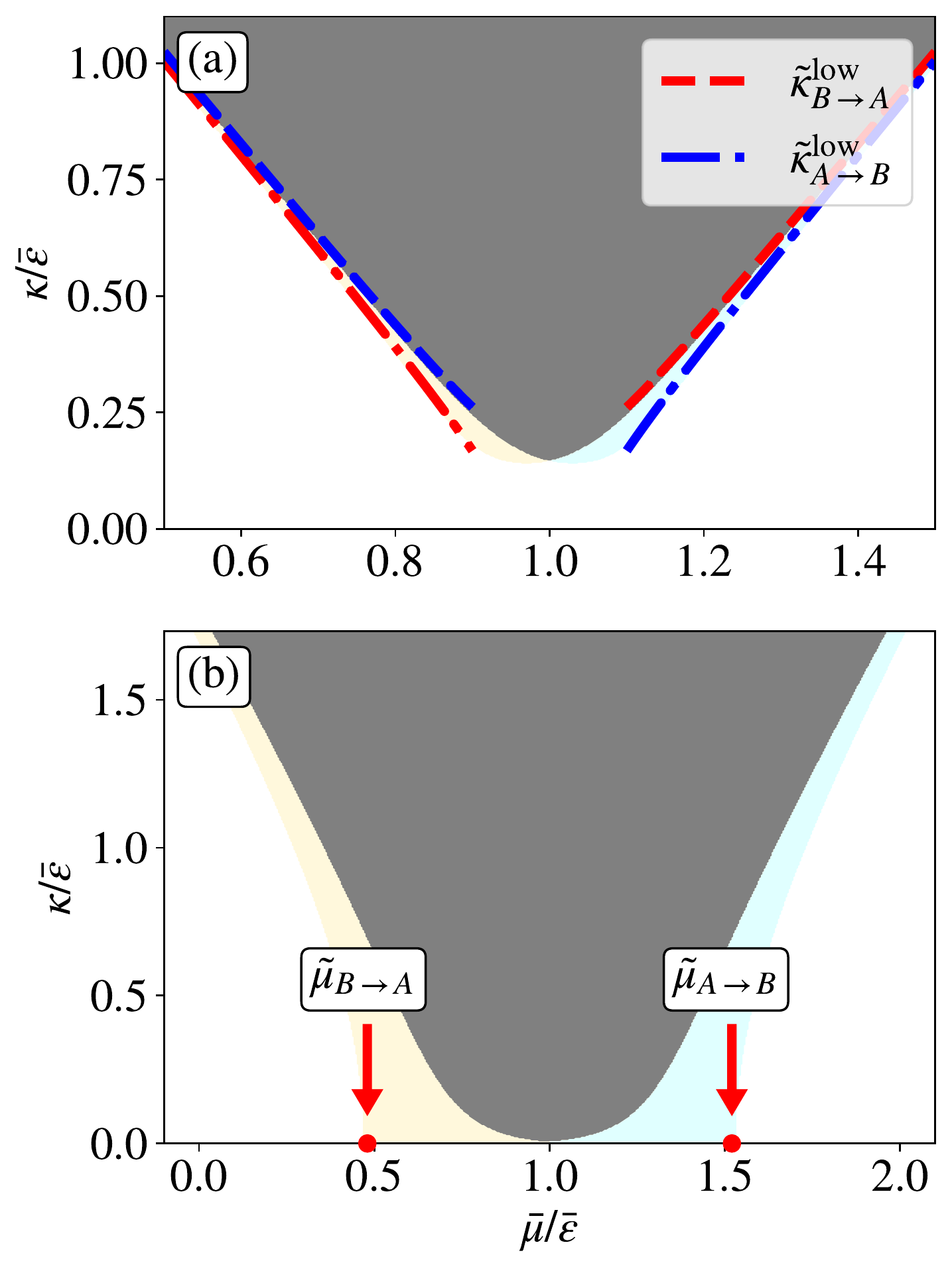}
    	\caption{Parameter regions of two-way steerable (grey), steerable from Bob to Alice (yellow) and steerable from Alice to Bob (blue). The threshold coupling strengths $\tilde \kappa^\text{low}_{A\rightarrow B}$ and $\tilde \kappa^\text{low}_{B\rightarrow A}$ are given by Eqs. (\ref{eq:kappa_f_A_B}) and (\ref{eq:kappa_f_B_A}). The detuning is set as (a) $\Delta\varepsilon=0.1\bar\varepsilon$; (b)(c) $\Delta\varepsilon=\bar\varepsilon$.}\label{fig k_mu_asy_eq}
    \end{center}
\end{figure}

As the two qubits are detuned, qubits $A$ and $B$ are not symmetric, implying the asymmetric steerability. At low temperature ($\beta|\bar\varepsilon-\bar\mu|\gg 0$) and small detuning $\Delta\varepsilon/\bar\varepsilon\ll 1$, we can derive the threshold coupling strength as
\begin{subequations}
\begin{align}
\label{eq:kappa_f_A_B}
    \tilde \kappa^\text{low}_{A\rightarrow B} = & \tilde \kappa^\text{low}_{A\leftrightarrow B} +\text{sgn}(\bar\varepsilon-\bar\mu) \frac{2\Delta\varepsilon}{\tilde \kappa^\text{low}_{A\leftrightarrow B}}T;\\
\label{eq:kappa_f_B_A}
    \tilde \kappa^\text{low}_{B\rightarrow A} = & \tilde \kappa^\text{low}_{A\leftrightarrow B} - \text{sgn}(\bar\varepsilon-\bar\mu) \frac{2\Delta\varepsilon}{\tilde \kappa^\text{low}_{A\leftrightarrow B}}T,
\end{align}
\end{subequations}
given by $\tilde\kappa^\text{low}_{A\leftrightarrow B}$ obtained from the symmetric qubit case. Steering from Alice to Bob is easier than the other direction if $\bar\mu>\bar\varepsilon$. When the symmetric $\tilde \kappa^\text{low}_{A\leftrightarrow B}$ is large, the asymmetric steerability vanishes. See Fig. \ref{fig k_mu_asy_eq} for the numerical results.  

Different from bosonic environments, the asymmetric steerability of the fermionic steady state comes from the population difference of the eigenstates $|00\rangle_{AB}$ and $|11\rangle_{AB}$. When $\bar\mu<\bar\varepsilon$, we have $p_{00}>p_{11}$ because of the lack of free electrons from environments. When $\bar\mu>\bar\varepsilon$, the full occupied state $|11\rangle_{AB}$ has a higher probability. Recall the expression of $f_b$ in Eq. (\ref{eq:f_b_detail}). The mismatch $p_{00}>p_{11}$ ($p_{00}<p_{11}$) gives positive (negative) of $f_b$, therefore giving asymmetric steerability from Bob to Alice (Alice to Bob). Note that we always have $p_->p_+$ because of the low temperature condition. In bosonic case, thermal excitations can not give higher populations of state $|11\rangle_{AB}$ compared to state $|00\rangle_{AB}$, because state $|11\rangle_{AB}$ always has a higher energy than state $|00\rangle_{AB}$. Therefore only asymmetric steerability from Bob to Alice is revealed (for $\varepsilon_A>\varepsilon_B$) in the bosonic cases.

We also find that, similar to the bosonic case shown in Eq. (\ref{eq:delta_w}), if the detuning is large enough ($\Delta\varepsilon/\bar\varepsilon  \gtrapprox 0.309$), there is a robust chemical potential $\tilde\mu_{A\rightarrow B}$ or $\tilde\mu_{B\rightarrow A}$. Below or above it, any nonzero coupling strength can give one direction of steering. See Fig. \ref{fig k_mu_asy_eq} for the illustrations.

\section{\label{sec:nonequilibrium} Steerability of nonequilibrium steady states}

\subsection{Heat/particle current and entropy production rate}

Nonequilibrium bosonic (fermionic) environments suggest the heat (particle) current flowing from the higher temperature (chemical potential) reservoir to the lower temperature (chemical potential) reservoir. Previous studies have revealed such heat or particle currents in the two-qubit setup \cite{QRRP07,WS11,WWW19,WW19,ZW20,ZWW20}. Heat or particle current can be characterized in terms of the energy or particle number change caused by the reservoir A or B. Specifically, the heat and particle current of reservoir $j=A,B$ is
\begin{equation}
    \label{eq:current}
    I^\text{b}_j = \tr(\mathcal D_j[\rho_{AB}]H_{AB}), \quad  I^\text{f}_j = \tr(\mathcal D_j[\rho_{AB}]\mathcal N_{AB}),
\end{equation}
with the dissipator $\mathcal D_j$ defined in Eq. (\ref{eq:master_eq2}). The superscript b or f distinguishes the bosonic or fermionic nonequilibrium environments. Note that the two-qubit system is treated as one combined system, therefore we can not distinguish the current flowing through (in) the qubit A or B. However, we can always say the current from the reservoir A or B since the two environments are well distinguished. We set the convention $I_j>0$, which means the energy or particle flowing from the reservoir $j$ to the system. 

In the steady state regime given by $\text{d}\rho_{AB}/\text{d}t = 0$, we have the continuity relation $I^\text{b}_A+I^\text{b}_B=0$ or $I^\text{f}_A+I^\text{f}_B=0$. The steady-state heat current is often linearly proportional to the temperature difference $\Delta T = T_B-T_A$, namely the Fourier's law \cite{WS08,LXLW17}. The steady-state particle current also often follows the linear increase in terms of the chemical potential difference $\Delta \mu = \mu_B-\mu_A$ in the near equilibrium regime. However, the particle current can saturate to a constant in the far from nonequilibrium regime due to the Pauli exclusion principle \cite{WWW19,WW19,ZW20,ZWW20}. 

The nonzero heat or particle current implies a nonzero entropy production of the system. If we assume that both the two reservoirs are infinitely large, therefore are approximately maintained at their equilibrium temperature or chemical potential. In a phenomenological way, the entropy production rate of the steady state can be characterized by
\begin{equation}
    \sigma^\text{b} = I^\text{b}_B \left(\frac{1}{T_A}-\frac{1}{T_B}\right),\quad \sigma^\text{f} =  I^\text{b}_B \left(\frac{\mu_B-\mu_A}{T}\right).
\end{equation}
We have assumed the same temperature for the two fermionic reservoirs. For $T_B>T_A$, we have $I_B^\text{b}>0$. For $\mu_B>\mu_A$, we have $I_B^\text{f}>0$. Therefore the positivity of $\sigma^\text{b}$ or $\sigma^\text{f}$ are guaranteed. The above definition of entropy production rate is taken from the classical case. Although quantum entropy production is different \cite{Landi20}, they stand for a good approximation when the environment is big. Temperature (chemical potential) difference can be viewed as the essential nonequilibrium condition (like an effective voltage or pressure); heat (particle) current is the corresponding nonequilibrium responses or behaviors; entropy production rate represents for the thermodynamic cost or power to maintain the nonequilibrium. In the following, we study the steerability with respect to these three nonequilibrium conditions.

\subsection{\label{subsec:b_nonequilibrium}Bosonic nonequilibrium environments}

\begin{figure}
    \begin{center}
    	\includegraphics[width=\columnwidth]{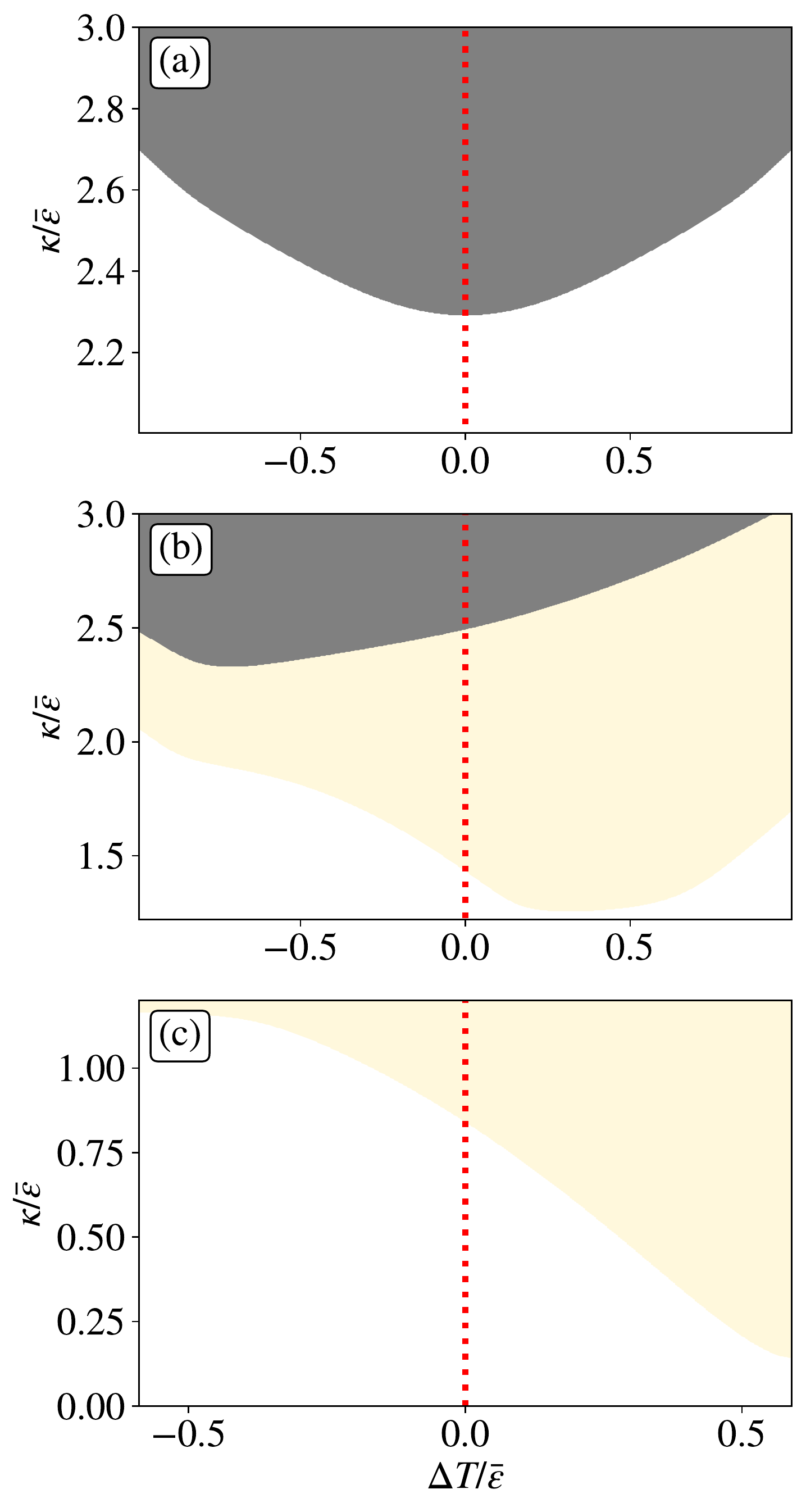}
    	\caption{Parameter regions of two-way steerable (grey) and steerable from Bob to Alice (yellow). The parameters are set as (a) $\Delta\varepsilon=0$ and $\bar T = 0.5\bar\varepsilon$; (b) $\Delta\varepsilon=1.6\bar\varepsilon$ and $\bar T = 0.5\bar\varepsilon$; (c) $\Delta\varepsilon=1.6\bar\varepsilon$ and $\bar T = 0.3\bar\varepsilon$. The range of $\kappa$ in (a)(b) is in the strong-coupling phase; in (c) is the weak-coupling phase. The coupling spectrum is set as $\gamma=0.01\bar\varepsilon$. The vertical dot line shows the equilibrium condition $\Delta T=0$.}\label{fig k_dT}
    \end{center}
\end{figure}

\begin{figure}
    \centering
    \includegraphics[width=\columnwidth]{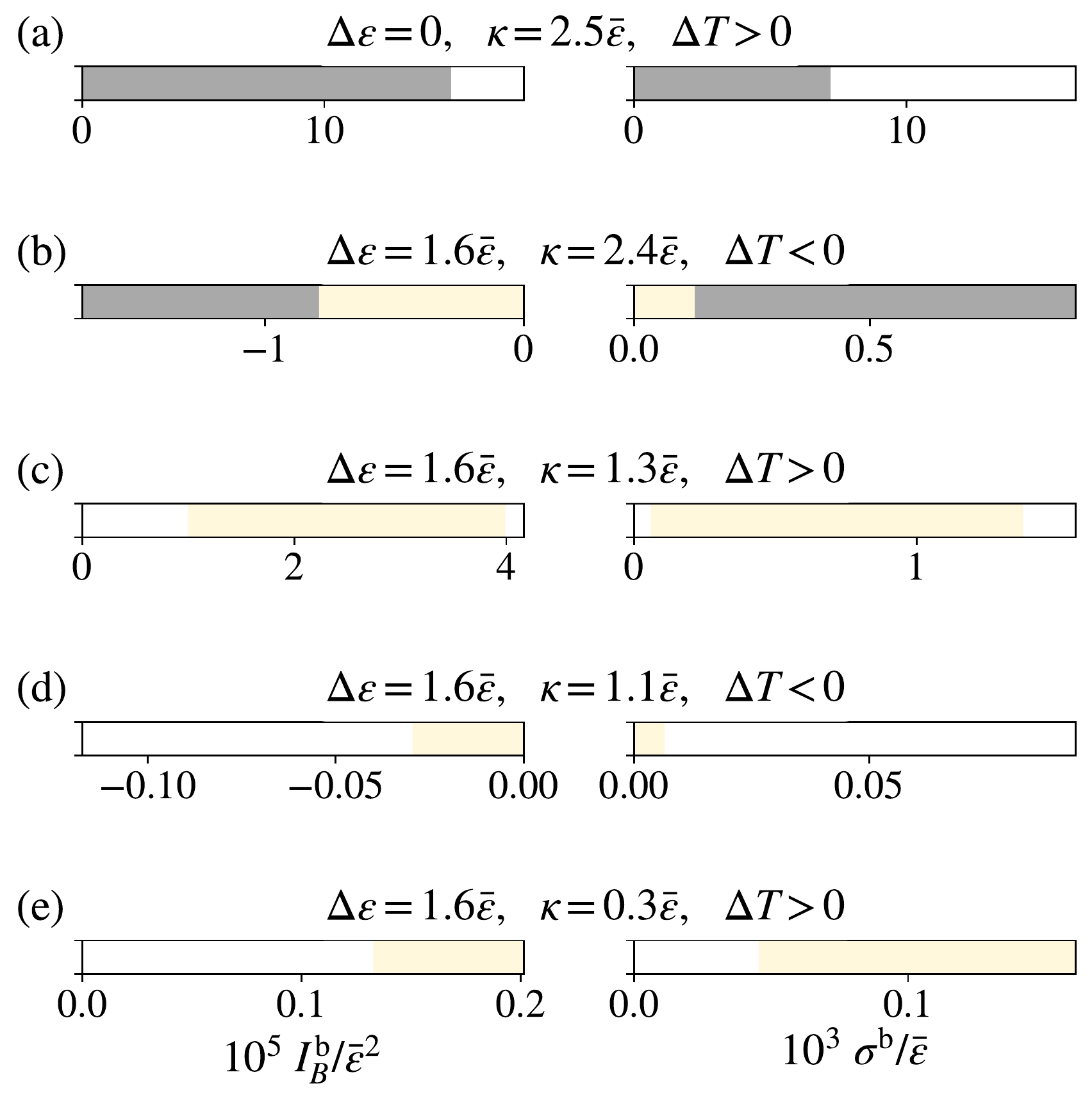}
    \caption{The one-dimensional plot of steerability in terms of heat current and entropy production rate. The yellow regions suggest the steerability from B to A. The grey regions suggest the two-way steerable. The mean temperature is fixed given by (a)(b)(c) $\bar T=0.5\bar\varepsilon$ and (d)(e) $\bar T=0.3\bar\varepsilon$. The coupling spectrum is set as $\gamma=0.01\bar\varepsilon$.}
    \label{fig:current_b}
\end{figure}

We consider how the nonequilibrium environment (with the fixed average temperature $\bar T = (T_A+T_B)/2$) affects the steerability of two qubits. Though the analytical solutions of nonequilibrium steady states are possible \cite{WWW19,ZWW20}, it is far too complicated to give the analytical analysis here. We discuss the numerical results instead. 

First, if the two qubits are symmetric, we find that the nonequilibrium environment does not generate the asymmetric steerability, even though the reduced density matrix is not symmetric in terms of $A$ and $B$. Note that we also do not have a steerable nonequilibrium steady state if the two qubits are symmetric and in the weak-coupling phase. Although the nonequilibrium environment contributes to the energy basis coherence \cite{WWW19,ZW14}, we do not have the population difference, such as $\rho_{22}\neq \rho_{33}$ for the symmetric qubits, in the nonequilibrium environment. See Fig. \ref{fig k_dT}. The necessary condition for asymmetric steerability is $\varepsilon_A\neq\varepsilon_B$. It is expected since only the asymmetric correlation gives the asymmetric steerability \cite{Bowles16,Yang2020}. Here we derive the Bloch-Redfield equation where the local environment collectively acts on the two-qubits, but the correlated eigenstate is always partially entangled if $\varepsilon_A\neq\varepsilon_B$. 

Second, if the two qubits are asymmetric and in the strong-coupling phase, we find that the nonequilibrium environment affects differently between the steerability from Bob to Alice and from Alice to Bob. As $\varepsilon_A>\varepsilon_B$, nonequilibrium environments with $T_B>T_A$ can enhance the steerability from Bob to Alice, while $T_A>T_B$ favors the steerability from Alice to Bob. We can intuitively understand that the nonequilibrium environments with $T_B>T_A$ adds more noises on Bob's state, therefore the steerability from Alice to Bob becomes weaker. One can also find that the nonequilibrium environments with $T_B>T_A$ will increase the population difference between eigenstates $|\psi^\pm(\theta)\rangle_{AB}$ given in Eqs. (\ref{eq:psi-}) and (\ref{eq:psi+}), which can enhance the steerability from Bob to Alice. Note that a higher probability of state $|\psi^-(\theta)\rangle_{AB}$ than state $|\psi^+(\theta)\rangle_{AB}$ means a larger factor $f_b$ with the expression in Eq. (\ref{eq:f_b_detail}), therefore it favors more the asymmetric steerability from Bob to Alice.

Previous study has shown that the nonequilibrium environments with $T_A>T_B$ can enhance the nonequilibrium steady-state entanglement and Bell nonlocality if $\varepsilon_A>\varepsilon_B$ \cite{ZW20}. In other words, qubit A with a larger gap is more robust to the thermal excitation compared to qubit B, corresponding to a higher quantum correlation because of a higher population of the ground entangled state. The enhanced two-way steerability due to the nonequilibrium environments with $T_A>T_B$ also follows a similar behavior. In general, stronger correlations suggest stronger two-way steerabilities. However, such rule is not guaranteed when we have the asymmetric steerability.

Third, we find that the steerability in the weak-coupling phase can also be enhanced by the nonequilibrium environments with $T_B>T_A$, similar to the asymmetric steerability in the strong-coupling phase. See Fig. \ref{fig k_dT}. We summarize the qualitative results for the influence of nonequilibrium bosonic environments on the steerability in Table \ref{tab:bosoinc}.

\begin{table*}[t]
    \centering
    \begin{tabular}{ccccc} 
     \hline\hline
    & Steerability from A to B & Steerability from A to B & Steerability from B to A & Steerability from B to A \\
    & with $T_A>T_B$ & with $T_B>T_A$ & with $T_A>T_B$ & with $T_B>T_A$ \\\hline
    $\kappa>2\sqrt{\varepsilon_A\varepsilon_B}$ & \multirow{2}*{Unfavorable} & \multirow{2}*{Unfavorable} & \multirow{2}*{Unfavorable} & \multirow{2}*{Unfavorable} \\
    with $\varepsilon_A=\varepsilon_B$ & && \\\hline
    $\kappa<2\sqrt{\varepsilon_A\varepsilon_B}$ & \multirow{2}*{NA} & \multirow{2}*{NA} & \multirow{2}*{NA} & \multirow{2}*{NA}\\
    with $\varepsilon_A=\varepsilon_B$ & && \\\hline
    $\kappa>2\sqrt{\varepsilon_A\varepsilon_B}$ & \multirow{2}*{Favorable} & \multirow{2}*{Unfavorable} & \multirow{2}*{Unfavorable} & \multirow{2}*{Favorable} \\
    with $\varepsilon_A>\varepsilon_B$ &&&&\\\hline
    $\kappa<2\sqrt{\varepsilon_A\varepsilon_B}$ & \multirow{2}*{NA} & \multirow{2}*{NA} & \multirow{2}*{Unfavorable} & \multirow{2}*{Favorable} \\
    with $\varepsilon_A>\varepsilon_B$ &&&&\\
\hline\hline
\end{tabular}
    \caption{Summary for the influence of nonequilibrium environments with $T_A\neq T_B$ on the steerability, based on the results shown in Fig. \ref{fig k_dT}. NA means no steering is found based on the entanglement-detection steering criterion.}
    \label{tab:bosoinc}
\end{table*}

We also study the steerability of the system in terms of the heat current and entropy production rate. Since the heat current and entropy production rate are also dependent on the inter-qubit coupling strength $\kappa$, we do not have the steerability regions like Fig. \ref{fig k_dT}. Instead, we numerically plot the one-dimensional value of the steerability in terms of the heat current and entropy production rate in Fig. \ref{fig:current_b}. For the symmetric qubits, Fig. \ref{fig:current_b} (a) shows that we always have the steerable to nonsteerable transition from equilibrium to nonequilibrium steady states, which corresponds to Fig. \ref{fig k_dT} (a). The symmetric qubits give symmetric currents and entropy production rates with respect to $|\Delta T|$. The symmetric qubits have nonequilibrium steady states which always have lower populations of the ground states compared to the equilibrium steady states, therefore may lose steerability in nonequilibrium environments.

On the other hand, we can see that the asymmetric qubits become two-way steerable (one-way steerable) from one-way steerable (nonsteerable), when we have nonzero heat current $I_B^\text{b}$ and positive entropy production rate. See Fig. \ref{fig:current_b} (b) and (c). It corresponds to Fig. \ref{fig k_dT} (b). Heat current is relatively blocked in one direction for asymmetric two qubits, called the thermal rectification \cite{SN05,WS11,ZW20}. In our case with $\varepsilon_A>\varepsilon_B$, the current from reservoir A to B is relatively blocked, while the system favors the steerability from B to A. In other words, the qubit A is harder to be thermal excited, and therefore blocks the heat current from reservoir A to B. This corresponds to the asymmetric steerability from B to A.

In the weak-coupling phase, if the coupling strength is comparable to $\bar\varepsilon$, the steerability from B to A vanishes for the nonequilibrium environments. However, the nonzero heat current and entropy production rate can be beneficial to the steerability from B to A if the coupling strength is relatively weak, such as $\kappa=0.3\bar\varepsilon$. See Fig. \ref{fig:current_b} (d) and (e), which corresponds to Fig. \ref{fig k_dT} (c). Existing a heat current suggests the connection between the two qubits, by which also implies the quantum correlation between the two qubits. However, when one environment has a much higher temperature (far from equilibrium), the correlation is classical and the steerability is also lost.

\subsection{\label{subsec:f_nonequilibrium}Fermionic nonequilibrium environments}

\begin{figure}
    \begin{center}
    	\includegraphics[width=\columnwidth]{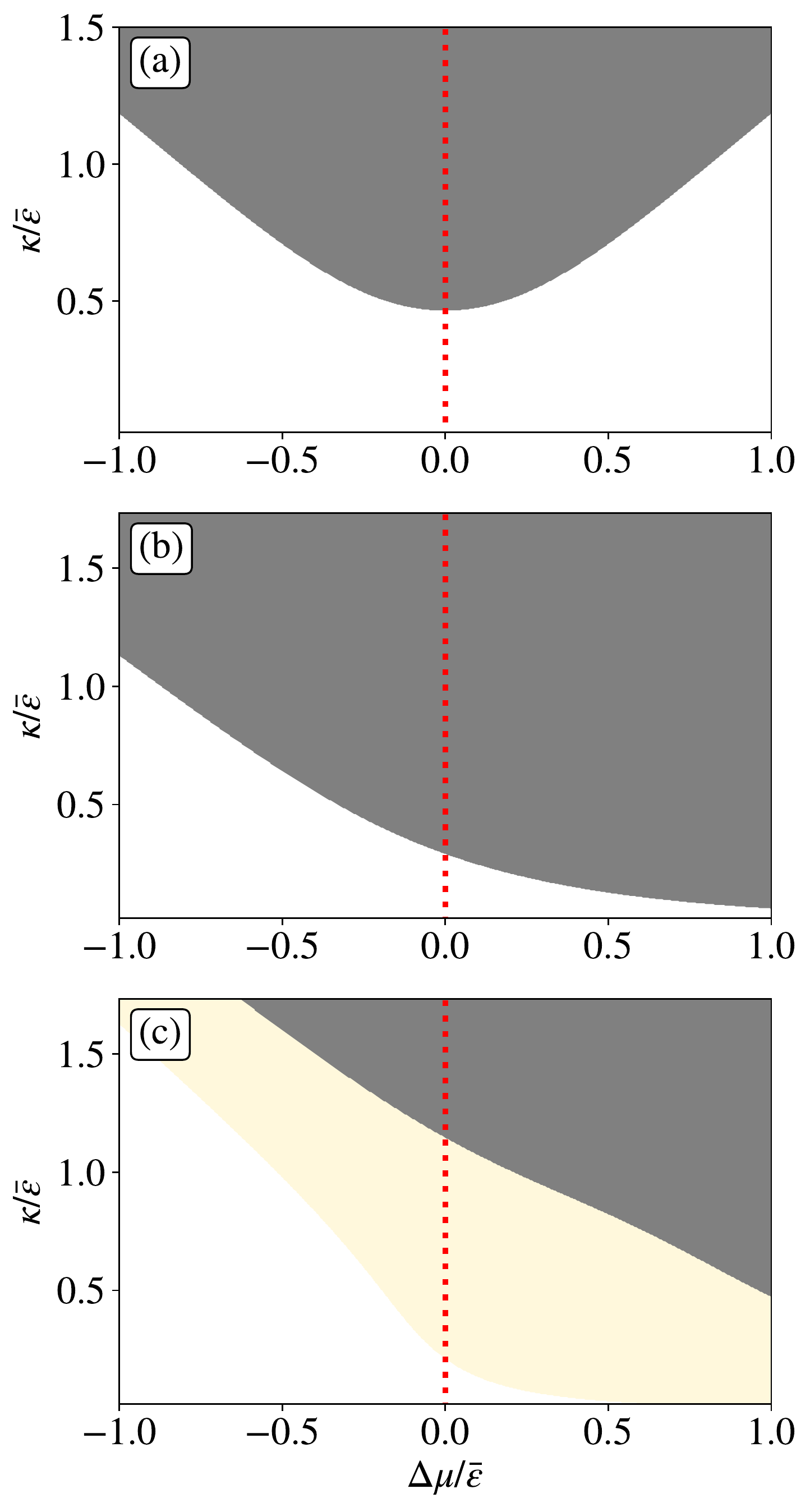}
    	\caption{Parameter regions of two-way steerable (grey) and steerable from Bob to Alice (yellow). The temperature is set as $\bar T = 0.15\bar\varepsilon$. The coupling spectrum is set as $\gamma=0.01\bar\varepsilon$. The other parameters are set as (a) $\Delta\varepsilon=0$ and $\bar \mu = \bar\varepsilon$; (b) $\Delta\varepsilon=\bar\varepsilon$ and $\bar \mu = \bar\varepsilon$; (c) $\Delta\varepsilon=\bar\varepsilon$ and $\bar \mu = 0.5\bar\varepsilon$. The vertical dot line shows the equilibrium condition $\Delta \mu=0$.}\label{fig k_dmu}
    \end{center}
\end{figure}

\begin{figure}
    \centering
    \includegraphics[width=\columnwidth]{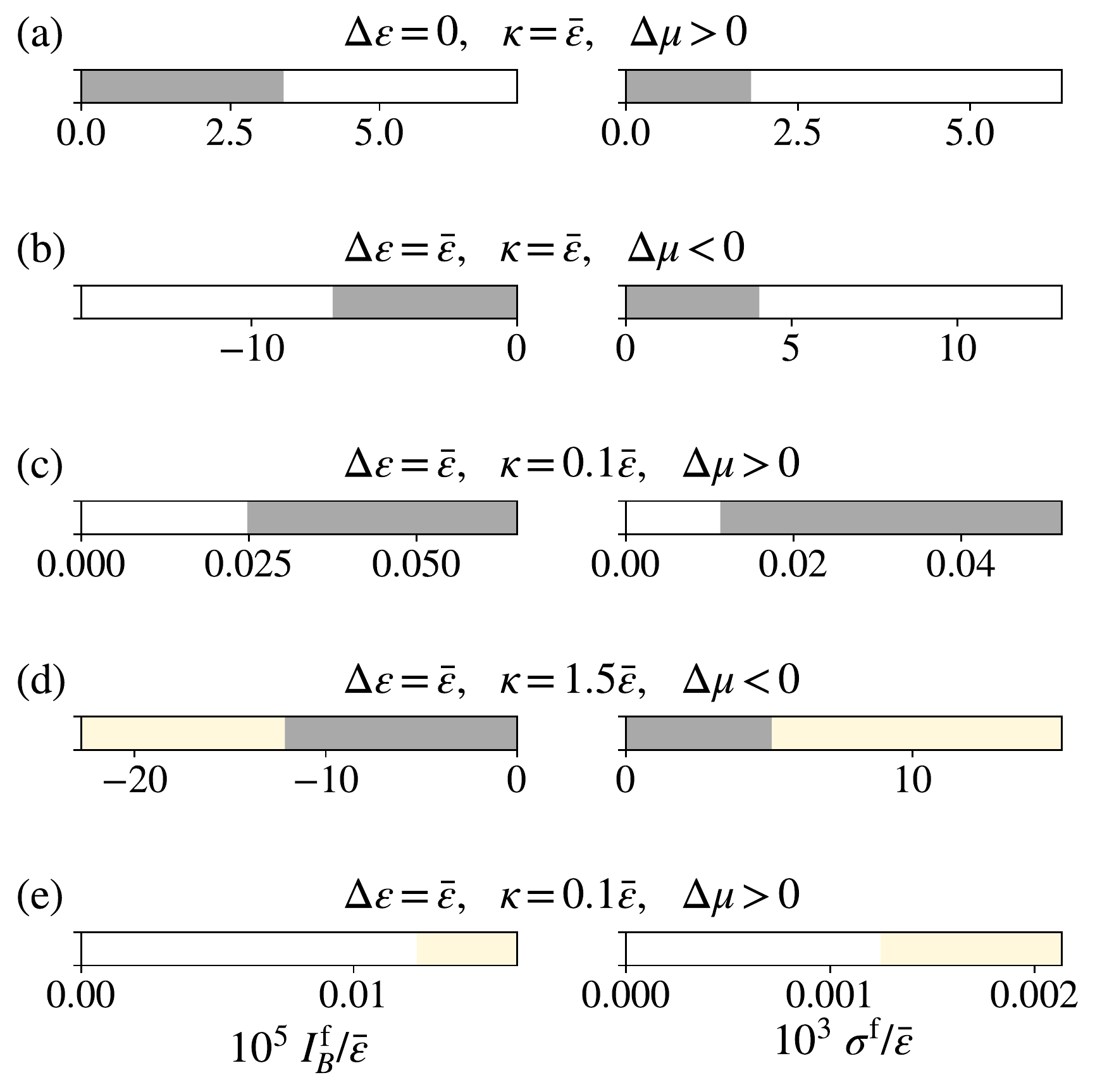}
    \caption{The one-dimensional plot of steerability in terms of particle currents and entropy production rates. The yellow regions suggest the steerability from B to A. The grey regions suggest the two-way steerable. The average chemical potential is (a)(b)(c) $\bar\mu = \bar\varepsilon$ and (d)(e) $\bar\mu = 0.5\bar\varepsilon$. The average temperature is $\bar T=0.15\bar\varepsilon$. The coupling spectrum is set as $\gamma=0.01\bar\varepsilon$.}
    \label{fig:current_f}
\end{figure}

To exclude the thermal effect which can lead to decoherence and classical descriptions, we study the fermionic nonequilibrium environments with $\mu_A\neq\mu_B$ (with the fixed average chemical potential $\bar\mu = (\mu_A+\mu_B)/2$) at the low temperature regime. Similar to the bosonic nonequilibrium environment, the fermionic nonequilibrium environment does not generate the asymmetric steerability with the symmetric two qubits. And the equilibrium environment requires the least coupling strength for steerability. See Fig. \ref{fig k_dmu} for numerical results. 

The asymmetric qubits give the asymmetric steerability in general. However, we find a hidden symmetry that the populations of states $|00\rangle_{AB}$ and $|11\rangle_{AB}$ are the same at the resonant point $\bar\mu = \bar\varepsilon$ even with the nonequilibrium environments. This corresponds to $\rho_{11} = \rho_{44}$, which gives $f_b=0$. Then the two steerability inequalities shown in Eqs. (\ref{eq:A_to_B}) and (\ref{eq:B_to_A}) are the same, which means no asymmetric steerability. Moreover, we find that qubit A with higher frequency coupled to lower chemical potential environment requires a weaker coupling between the two qubits for steerability. See Fig. \ref{fig k_dmu}. It has the similar behaviors in terms of entanglement and Bell nonlocality of the nonequilibrium steady states \cite{WWW19,ZW20}. 

In the equilibrium case, we have shown that the condition $\bar\varepsilon>\bar\mu$ ($\bar\varepsilon<\bar\mu$) favors the steerability from Bob to Alice (Alice to Bob) in terms of the asymmetric qubits with $\varepsilon_A>\varepsilon_B$. Numerically, we find that the nonequilibrium fermionic environment influences the two directions of steerability in a similar way: a higher frequency qubit coupled to a lower chemical potential environment requires a smaller coupling strength for steerability. See Fig. \ref{fig k_dmu}. In weak-coupling phase, the steady-state steerability comes from the excited state which is entangled. Then it is expected that the qubit with a lower energy gap coupled to the environment with a higher chemical potential can lead to more populations of excited entangled state, therefore a stronger quantum correlation as well as steerability. We summarize the qualitative results for the influence of nonequilibrium fermionic environments on the steerability in Table \ref{tab:fermionic}.

Corresponding to Fig. \ref{fig k_dmu} (a), Fig. \ref{fig:current_f} (a) shows that we do not have the steerability generated from the nonequilibrium environments with nonzero particle currents and entropy production rates. Fig. \ref{fig:current_f} (b) shows that the nonequilibrium steady states of asymmetric two qubits lose the steerability with $\kappa=\bar\varepsilon$. However, Fig. \ref{fig:current_f} (c) reveals that the asymmetric two qubits have the steerable nonequilibrium steady states, contrast to the equilibrium steady states with zero particle currents and entropy production rates. Note that the current requires the particle excitation in one qubit and relaxation in another qubit. Quantum correlation is generated in such dynamical process, therefore the steerability may also be favored. We do not have the particle current rectification effect here, which suggests the same responses of steerability in two directions with respect to the same particle currents. The results of Figs. \ref{fig:current_f} (b) and (c) are consistent with the results given by Fig. \ref{fig k_dmu} (b).

When the coupling strength $\kappa$ is comparable to $\bar\varepsilon$ (but still in the weak-coupling phase), we have the two-way steerable equilibrium steady states while the nonequilibrium steady states might be only steerable in one direction. See Fig. \ref{fig:current_f} (d). When the coupling strength $\kappa$ is weak, such as $\kappa = 0.1\bar\varepsilon$, we have the steerable states given by the nonequilibrium environments with positive particle currents and entropy production rates. See Fig. \ref{fig:current_f} (e). The results given by Figs. \ref{fig:current_f} (d) and (e) are qualitatively the same as the results in Fig. \ref{fig k_dmu} (c). Similar with the bosonic cases, the steerability can come from the correlation related to the particle current at small $\kappa$. But such correlations can be fragile especially for weak coupled two qubits, and a large current does not support steerability in general.

\begin{table*}[t]
    \centering
    \begin{tabular}{ccccc} 
    \hline\hline
    & Steerability from A to B & Steerability from A to B & Steerability from B to A & Steerability from B to A \\
    & with $\mu_A>\mu_B$ & with $\mu_B>\mu_A$ & with $\mu_A>\mu_B$ & with $\mu_B>\mu_A$ \\\hline
    $\kappa<2\sqrt{\varepsilon_A\varepsilon_B}$ & \multirow{2}*{Unfavorable} & \multirow{2}*{Unfavorable} & \multirow{2}*{Unfavorable} & \multirow{2}*{Unfavorable} \\
    with $\varepsilon_A=\varepsilon_B$ & && \\\hline
    $\kappa<2\sqrt{\varepsilon_A\varepsilon_B}$ & \multirow{2}*{Unfavorable} & \multirow{2}*{Favorable} & \multirow{2}*{Unfavorable} & \multirow{2}*{Favorable}\\
    with $\varepsilon_A>\varepsilon_B$ & && \\
    \hline\hline
    \end{tabular}
    \caption{Summary for the influence of nonequilibrium environments with $\mu_A\neq\mu_B$ on the steerability, based on the results shown in Fig. \ref{fig k_dmu}. The temperatures of the two reservoirs are set as equal $\bar T = T_A=T_B$, and in the low temperature regimes $\bar T\ll\bar\varepsilon$. }
    \label{tab:fermionic}
\end{table*}

\section{\label{sec:comparison} Relationship between entanglement, steerability and Bell nonlocality}

\subsection{Entanglement and Bell nonlocality}

Entanglement and Bell nonlocality are other two types of quantum correlations. Two-qubit states without a local factorized formalism are called entangled \cite{Horodecki09}. The two-qubit entangled state as well as the entanglement criteria are well studied. Applying the positive partial transpose criterion on the ``X''-state given in Eq. (\ref{def:X_state}) gives the entanglement inequality 
\begin{equation}
    |\rho_{23}|^2>\rho_{11}\rho_{44},
\end{equation}
assuming $|\rho_{23}|\gg |\rho_{14}|$ which is valid in our two-qubit model. It is easy to see that any state satisfies the steering inequality shown in Eqs. (\ref{eq:A_to_B}) or (\ref{eq:B_to_A}), also satisfies the above entanglement inequality. It also follows the entanglement-detection steering criterion, since the steering inequalities are given by the entanglement criterion on the state with local noises. 

Bell nonlocality describes that the statistics of local measurements on the correlated states can not be descried by any local hidden variable theory \cite{Brunner14}. It is surprising to see that not all entangled states are Bell nonlocal \cite{Werner89}. Based on the CHSH inequality \cite{CHSH69}, there is a simple necessary and sufficient criterion to judge Bell nonlocality \cite{HHH95}. In the ``X''-state setup, states satisfying
\begin{equation}
\label{eq:Bell_ineq}
    |\rho_{23}|^2>\frac 1 8,\quad \text{or} \quad |\rho_{23}|^2>\frac 1 4-\frac 1 4\left(2(\rho_{22}+\rho_{33})-1\right)^2
\end{equation}
can violate the CHSH inequality. Note that both entanglement, steerability and Bell nonlocality inequalities have the formalism that the local state coherence $\rho_{23}$ is larger than a certain threshold given by the local state population terms.

\subsection{Hierarchy of correlation in bosonic setup}

\begin{figure}
    \begin{center}
    	\includegraphics[width=\columnwidth]{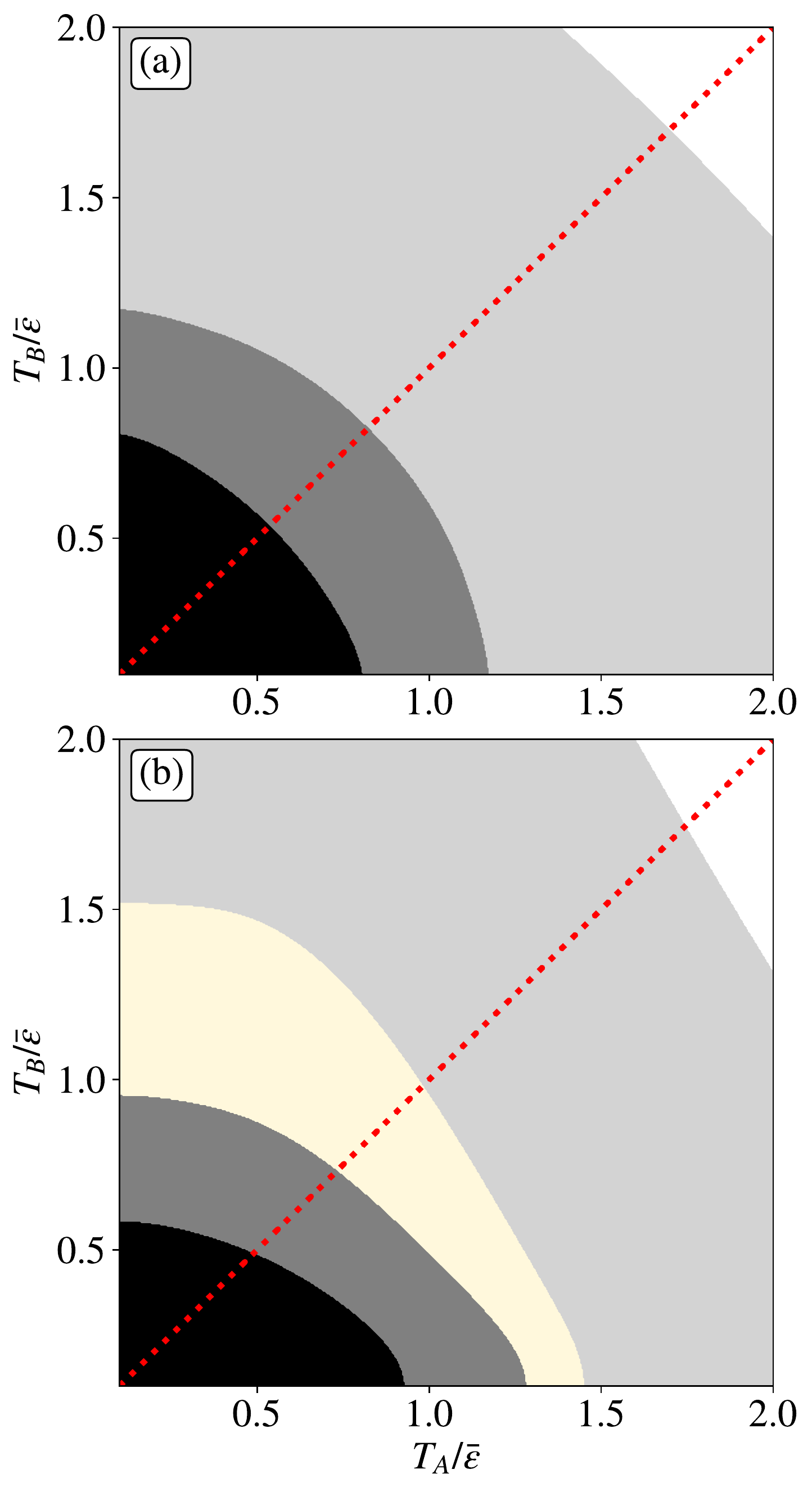}
    	\caption{Parameter regions of Bell nonlocality (black), two-way steerability (dark grey), steerable from Bob to Alice (yellow), and entanglement (light grey). The parameters are set as $\kappa = 3\bar\varepsilon$, $\gamma = 0.01\bar\varepsilon$, (a) $\Delta\varepsilon = 0$, and (b) $\Delta\varepsilon = 1.6\bar\varepsilon$. The dot red line represents for $T_A=T_B$.}\label{fig contour_b}
    \end{center}
\end{figure}

In the equilibrium setup, we have shown that the symmetric two qubits do not have the steerability, and the threshold coupling strength $\tilde \kappa^\text{low}_{A\leftrightarrow B}$ follows Eq. (\ref{eq:kappa_low}). Previous studies have shown that the equilibrium steady state can be entangled in the weak-coupling phase \cite{QRRP07,WW19,WWW19}, and the threshold coupling strength is
\begin{equation}
\label{eq:kappa_ent}
    \tilde\kappa_\text{Ent} = 2\ln\left(1+\sqrt 2\right)T.
\end{equation}
On the other hand, based on inequality (\ref{eq:Bell_ineq}), we find the threshold coupling strength in terms of Bell nonlocality at low temperature is
\begin{equation}
    \tilde\kappa^\text{low}_\text{Bell} = 2\bar\varepsilon+ 2\ln\left(\frac{1}{\sqrt 2-1}\right)T,
\end{equation}
which clearly shows that Bell nonlocality can only exist in the strong-coupling phase. Comparing to the threshold coupling strength $\tilde \kappa^\text{low}_{A\leftrightarrow B}$ for steerability, we have the hierarchy given by
\begin{equation}
    \tilde\kappa^\text{low}_\text{Bell} > \tilde \kappa^\text{low}_{A\leftrightarrow B} > \tilde\kappa_\text{Ent}.
\end{equation}
This follows the intuitive understanding, since a strong correlation requires a strong coupling between the two qubits. 

As for nonequilibrium steady states, we numerically verify the hierarchy of correlations in Fig. \ref{fig contour_b}. For symmetric qubits, entanglement, steerability and Bell nonlocality all show the symmetries respect to $T_A$ and $T_B$. It is also interesting to see that the boundary separating the entangled and unentangled states is almost a straight line in terms of $T_A$ and $T_B$. Based on the analytical results on the equilibrium environment, we can conjecture that the two symmetric qubits with the nonequilibrium temperature (assuming that $\Delta T\ll \bar T$) satisfying 
\begin{equation}
    T_A + T_B < \frac{2\kappa}{\ln(1+\sqrt 2)},
\end{equation}
are entangled. Note that the threshold coupling strength for entanglement, defined in Eq. (\ref{eq:kappa_ent}) is also valid in the relative high temperature regime. 

For asymmetric two qubits $\varepsilon_A>\varepsilon_B$, Fig. \ref{fig contour_b} shows that the threshold temperature of environment $A$ (the intercept point on $x$ axis) for the two-way steerability is larger than the threshold temperature of environment $B$ (the intercept point on $y$ axis). The above statement is consistent with the result on how the nonequilibrium environment influences the steerability in two different directions, discussed in Sec. \ref{subsec:b_nonequilibrium}. The steerability from Bob to Alice comes from the population difference of the partially entangled states $|\psi^\pm(\theta\rangle$ defined in Eqs. (\ref{eq:psi-}) and (\ref{eq:psi+}). On the other hand, stronger correlations such as entanglement and Bell nonlocality suggest the stronger two-way steerability.

\subsection{Hierarchy of correlation in fermionic setup}

\begin{figure}
    \begin{center}
    	\includegraphics[width=\columnwidth]{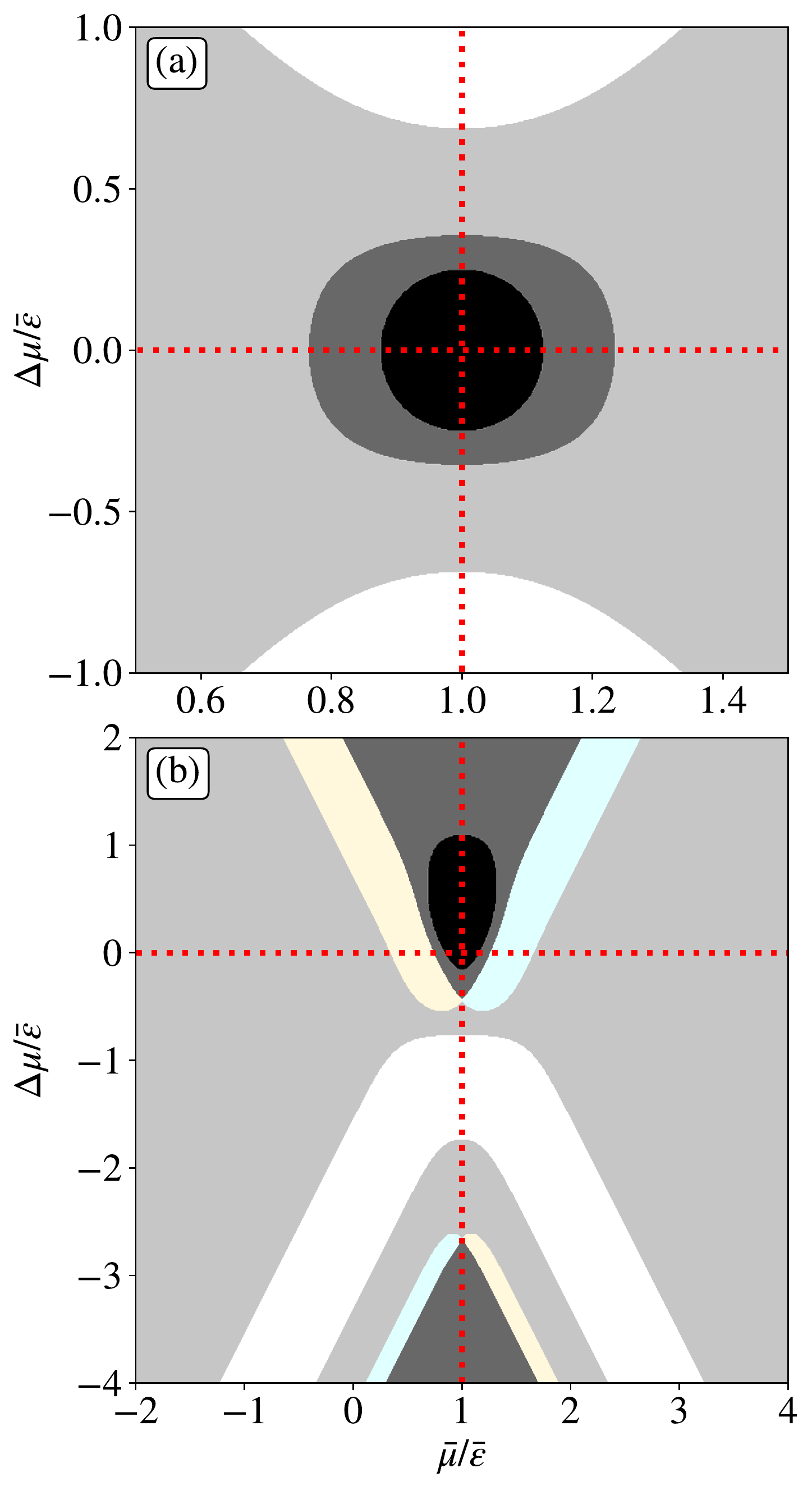}
    	\caption{Parameter regions of Bell nonlocality (black), two-way steerability (dark grey), steerable from Bob to Alice (yellow), steerable from Alice to Bob (blue), and entanglement (light grey). The parameters are set as $\kappa = 0.6\bar\varepsilon$, $\gamma = 0.01\bar\varepsilon$, $\bar T=0.15\varepsilon$, (a) $\Delta\varepsilon = 0$, and (b) $\Delta\varepsilon = \varepsilon$. The horizontal and vertical dot red lines represent for $\Delta\mu=0$ and $\bar\mu = \bar\varepsilon$ respectively.}\label{fig contour_f}
    \end{center}
\end{figure}

For fermionic environments, we also study the hierarchy of these three different correlations. At equilibrium, two symmetric qubits are entangled if $\kappa>\tilde\kappa^\text{low}_\text{Ent}$ with \cite{WW19,WWW19}
\begin{equation}
\label{eq:kappa_f_ent}
    \tilde\kappa_\text{Ent} = 2\ln\left(1+\sqrt 2\right)T,
\end{equation}
which is identical to the bosonic case given by Eq. (\ref{eq:kappa_ent}). Note that the threshold coupling strength is independent of the equilibrium chemical potential. Bell nonlocality requires a stronger coupling strength between the two qubits. At the resonant point $\bar\mu = \bar\varepsilon$, we find the threshold coupling strength 
\begin{equation}
    \tilde\kappa_\text{Bell}(\bar\mu = \bar\varepsilon) = 2\ln\left(3+2\sqrt 2\right)T.
\end{equation}
Away from the resonant point, the threshold coupling strength for Bell nonlocality is
\begin{equation}
\label{eq:kappa_f_away_bell}
    \tilde\kappa^\text{low}_\text{Bell} = 2|\bar\varepsilon-\bar\mu|+2\ln\left(\frac{1}{\sqrt 2-1}\right)T,
\end{equation}
if $\beta|\bar\varepsilon-\bar\mu|\gg 0$ (low temperature). Compare to the threshold coupling strength for steerability given in Eqs. (\ref{eq:kappa_mu_w}) and (\ref{eq:kappa_f_away}), we have the hierarchy at the resonant point
\begin{equation}
    \tilde\kappa_\text{Bell}(\bar\mu = \bar\varepsilon) > \tilde\kappa_{A\leftrightarrow B}(\mu = \bar\varepsilon) > \tilde\kappa_\text{Ent},
\end{equation}
and away from the resonant point
\begin{equation}
    \tilde\kappa^\text{low}_\text{Bell} > \tilde\kappa^\text{low}_{A\leftrightarrow B} > \tilde\kappa_\text{Ent}.
\end{equation}
Both steerability and Bell nonlocality not only require the near resonant condition $\bar\varepsilon\approx \bar\mu$, but also a relative low temperature. However, entanglement can survive for environments with any chemical potential as long as the temperature is low. 

We plot the hierarchy of the nonequilibrium steady-state correlations in Fig. \ref{fig contour_f}. For symmetric two qubits, we can see that entanglement, steerability and Bell nonlocality are all symmetric with respect to the lines $\Delta\mu=0$ and $\bar\mu = \bar\varepsilon$. The equilibrium at the resonant point gives the strongest correlation. 

The nonequilibrium steady states of the two asymmetric qubits have richer structures. First, in the near equilibrium regime, if the average chemical potential $\bar\mu$ is below (above) $\bar\varepsilon$, we have a higher (lower) population of state $|00\rangle_{AB}$ than state $|11\rangle_{AB}$, therefore positive (negative) of parameter $f_b$ defined in Eq. (\ref{eq:f_b}), which suggests the asymmetric steerability from Bob to Alice (Alice to Bob). In other words, the direction of asymmetric steerability is dominated by the average chemical potential of the two reservoirs. This is consistent with the results in Sec. \ref{subsec:f_nonequilibrium}.

%The asymmetric steerability comes from the population difference of the partial entangled eigenstate $\psi^\pm\rangle_{AB}$ denoted in Eqs. (\ref{eq:psi-}) and (\ref{eq:psi+}). More specifically, higher (lower) population of state $\psi^+\rangle_{AB}$ ($\psi^-\rangle_{AB}$) means asymmetric steerability from Alice to Bob (Bob to Alice).

Second, the above asymmetric steerability reverses the direction in the far from equilibrium cases $\mu_A\gg \mu_B$. In the near equilibrium regime, we always have $\rho_{22}>\rho_{33}$ because of $\varepsilon_A>\varepsilon_B$. In far from equilibrium cases $\mu_A\gg \mu_B$, the chemical potential gives the effective energies of two qubits $\varepsilon'_A$ and $\varepsilon'_B$ which leads to $\varepsilon'_A<\varepsilon'_B$. Then the relationship between the populations of local states become $\rho_{22}<\rho_{33}$. Therefore the asymmetric steerability is also reversed. Note that the resonance condition $\bar\varepsilon=\bar\mu$ always suppresses the asymmetric steerability for asymmetric two qubits, because $\rho_{11} = \rho_{44}$ gives $f_b=0$ defined in Eq. (\ref{eq:f_b}).

%One can check that the asymmetric steerability from Bob to Alice (Alice to Bob) emerges when the population of eigenstate $|\psi^-\rangle$ (first-excited state) is more (less) than the eigenstate $|\psi^+\rangle$ (second-excited state). Third, the far from equilibrium correlations are relatively weak, and we do not observe Bell nonlocality in such regime.

%The correlations generated in such far from equilibrium cases are due to the coherence contributed from the nonequilibrium environment \cite{WW19,WWW19}. 

\section{\label{sec:conclusion}Conclusions}

We study the steerability of two interacting qubits, which couple to equilibrium or nonequilibrium environments. Under bosonic environments, we find that the detuned qubits with $\varepsilon_A>\varepsilon_B$ favors the steerability from Bob to Alice. Nonequilibrium environments with $T_B>T_A$, giving positive entropy production rates, can enhance such asymmetric steerability from Bob to Alice, while suppress the steerability in the other direction. Under fermionic environments, we find that asymmetric steerability from Alice to Bob (Bob to Alice) emerges if the chemical potential is above (below) the resonant value $\bar\varepsilon$. Nonequilibrium environments with different chemical potentials with positive entropy production rates can enhance the steerability if the two qubits are detuned. We obtain the threshold coupling strength for entanglement, steerability and Bell nonlocality, and quantify their hierarchy. Our results can be verified on the superconducting qubits \cite{Pashkin03} and quantum dot qubits \cite{Xiang13} for the bosonic and fermionic setups respectively. Our study is helpful for designing different quantum information processing tasks where different quantum resources are required.

\begin{acknowledgments}

The authors thank Xuanhua Wang for helpful discussions.

\end{acknowledgments}

\appendix

\section{\label{App:A} Evolution matrix}

\subsection{Weak-coupling phase}

In the weak-coupling phase, the ground state and the first excited state have the energies $-\bar\varepsilon$ and $-\Omega$ respectively. Here $\bar\varepsilon$ is the average frequency of the two qubits, given by $\bar\varepsilon = (\varepsilon_{A}+\varepsilon_{B})/2$ and $\Omega = \sqrt{\Delta \varepsilon+\kappa^2}$ with $\Delta \varepsilon = \varepsilon_{A}-\varepsilon_{B}$. The second and third excited states have the energies $\Omega$ and $\bar\varepsilon$ respectively. We denote the four eigenstates of the two qubits as $\{|g\rangle,|e_1\rangle,|e_2\rangle,|e_3\rangle\}$. There are only two types of transitions, limited by the spin-boson model in Eq. (\ref{def:H_I}). Specifically, we have the transitions $g\leftrightarrow e_1$ and $e_2 \leftrightarrow e_3$ with the energy change $\varepsilon_- = \bar\varepsilon-\Omega$, and the transitions $g\leftrightarrow e_2$ and $e_1 \leftrightarrow e_3$ with the energy change $\varepsilon_+ = \bar\varepsilon+\Omega$.

Since we have assumed the coupling strength between the system and environment is much weaker than the coupling strength between two qubits, the two qubits are treated as one system. Therefore, it is more natural to work on the energy basis of the system. In the energy basis, the steady state density matrix has six nonvanishing entries. We can write it into a vector, denoted as
$$
|\rho^\text{ss}_{S}\rangle = \left(\rho_{gg},\rho_{e_1e_1},\rho_{e_2e_2},\rho_{e_3e_3},\rho_{e_1e_2},\rho_{e_2e_1}\right)^T,
$$
with matrix transpose $T$ (not the temperature). Then the evolution matrix $\mathcal M$ defined in Eq. (\ref{def:M}) is a $6\times6$ matrix. To simplify the matrix formalism, we define the parameters
\begin{subequations}
\begin{align}
\label{def_p_q1}
    &p =  \gamma\cos^2(\theta/2) n_{A}(\varepsilon_+) + \gamma\sin^2(\theta/2) n_{B}(\varepsilon_+);\\
\label{def_p_q3}
    &q =  \gamma\sin^2(\theta/2) n_{A}(\varepsilon_-)+\gamma\cos^2(\theta/2) n_{B}(\varepsilon_-);\\
    &s_{j} =  \frac \gamma 2 \sin\theta\left(n_{j}(\varepsilon_+)+n_{j}(\varepsilon_-)\right);\\
\label{def_p_q2}
    &t_{j} =  \frac \gamma 2 \sin\theta\left(n_{j}(\varepsilon_+)-n_{j}(\varepsilon_-)\right),
\end{align}
\end{subequations}
with $j=A,B$. Here $n_j$ is the distribution of the reservoir $j$, given by
\begin{equation}
    n_j(\varepsilon)=\frac{1}{\exp\left((\varepsilon-\mu_j)/T_j\right)\pm1}.
\end{equation}
The negative/positive sign is for the bosonic/fermionic distribution. 

The bosonic and fermionic setup gives the evolution matrices
\begin{widetext}
\begin{subequations}
\begin{equation}
    \mathcal M_\text{weak}^\text{b} = \left(
    \begin{array}{cccccc}
    -2(p+q) & 2(\gamma+q) & 2(\gamma+p) & 0 & s_B-s_A & s_B-s_A \\
    2q & -2(\gamma+p+q) & 0 & 2(\gamma+p) & t_A-t_B & t_A-t_B \\ 
    2p & 0 & -2(\gamma+p+q) & 2(\gamma+q) & t_B-t_A & t_B-t_A \\
    0 & 2p & 2q & -2(2\gamma+p+q) & s_A-s_B & s_A-s_B \\
    s_B-s_A & t_B-t_A & t_A-t_B & s_A-s_B & i\Omega-2(\gamma+p+q) & 0 \\
    s_B-s_A & t_B-t_A & t_A-t_B & s_A-s_B & 0 & -i\Omega-2(\gamma+p+q)
    \end{array} 
    \right); 
\end{equation}
\begin{equation}
    \mathcal M_\text{weak}^\text{f} = \left(
    \begin{array}{cccccc}
    -2(p+q) & 2(\gamma-q) & 2(\gamma-p) & 0 & s_A-s_B & s_A-s_B \\
    2q & -2(\gamma+p-q) & 0 & 2(\gamma-p) & s_B-s_A & s_B-s_A \\ 
    2p & 0 & -2(\gamma-p+q) & 2(\gamma-q) & s_B-s_A & s_B-s_A \\
    0 & 2p & 2q & -2(2\gamma-p-q) & s_A-s_B & s_A-s_B \\
    s_B-s_A & s_B-s_A & s_B-s_A & s_B-s_A & i\Omega-2\gamma & 0 \\
    s_B-s_A & s_B-s_A & s_B-s_A & s_B-s_A & 0 & -i\Omega-2\gamma
    \end{array} 
    \right).
\end{equation}
\end{subequations}
\end{widetext}
With the equilibrium environments, we have $s_A=s_B$ and $t_A=t_B$. Therefore the evolution matrices are block diagonal in terms of the population and coherence subspace. Then there is no steady-state coherence at the equilibrium environments. However, the evolution matrix at the nonequilibrium case gives nonzero steady-state coherence. Note that the Lindblad always decouples the population and coherence subspaces, therefore no steady-state coherence in any case.  

\subsection{Strong-coupling phase}

In the strong-coupling phase, the ground state becomes the singlet state. And the transitions $g\leftrightarrow e_1$ and $e_2 \leftrightarrow e_3$ exchange energy $\varepsilon_- = \Omega-\bar\varepsilon$; the transitions $g\leftrightarrow e_2$ and $e_1 \leftrightarrow e_3$ has the energy change $\varepsilon_+ = \bar\varepsilon+\Omega$.

We only consider the bosonic setup in the strong-coupling phase, since the electronic tunnelling is weak. We apply the same parameters defined in Eqs. (\ref{def_p_q1})-(\ref{def_p_q2}). Then the evolution matrix is
\begin{widetext}
\small
\begin{equation}
    \mathcal M_\text{strong}^\text{b} = \left(
    \begin{array}{cccccc}
    -2(p+q) & 2(\gamma+q) & 2(\gamma+p) & 0 & -s_A-s_B-2\gamma\sin\theta & -s_A-s_B-2\gamma\sin\theta \\
    2q & -2(\gamma+p+q) & 0 & 2(\gamma+p) & t_A+t_B+\gamma\sin\theta & t_A+t_B+\gamma\sin\theta \\ 
    2p & 0 & -2(\gamma+p+q) & 2(\gamma+q) & -t_A-t_B+\gamma\sin\theta & -t_A-t_B+\gamma\sin\theta \\
    0 & 2p & 2q & -2(2\gamma+p+q) & s_A+s_B & s_A+s_B \\
    -s_A-s_B & -t_A-t_B+\gamma\sin\theta & t_A+t_B+\gamma\sin\theta & s_A+s_B+2\gamma\sin\theta & i\Omega-2(\gamma+p+q) & 0 \\
    -s_A-s_B & -t_A-t_B+\gamma\sin\theta & t_A+t_B+\gamma\sin\theta & s_A+s_B+2\gamma\sin\theta & 0 & -i\Omega-2(\gamma+p+q)
    \end{array} 
    \right).
\end{equation}
\end{widetext}
The block matrix in the population subspace of the evolution matrices $\mathcal M_\text{weak}^\text{b}$ and $\mathcal M_\text{strong}^\text{b}$ are the same. It suggests that the weak- and strong-coupling phases have the same steady-state population in the leading order. However, the transformation matrices between the energy and local bases are different in these two phases, which gives different steady states in the local basis. 

\section{\label{App:B} Equilibrium steady state}

In the bosonic case, we set the chemical potential as zero. In the equilibrium setup, the parameters $p$ and $q$ defined in Eqs. (\ref{def_p_q1}) and (\ref{def_p_q3}) can be simplified as
\begin{equation}
    p = \frac{1}{e^{\beta\varepsilon_+}-1},\quad q = \frac{1}{e^{\beta\varepsilon_-}-1},
\end{equation}
with $\beta = \beta_A=\beta_B$. In the energy basis ($\{|g\rangle,|e_1\rangle,|e_2\rangle,|e_3\rangle\}$), we have the steady state solution
\begin{subequations}
\begin{align}
    &\rho_{gg} = \frac{1}{R}(1+p)(1+q);\\
    &\rho_{e_1e_1} = \frac{1}{R}(1+p)q;\\
    &\rho_{e_2e_2} = \frac{1}{R}(1+q)p;\\
    &\rho_{e_3e_3} = \frac{pq}{R},
\end{align}
\end{subequations}
with the normalization $R = (1+2p)(1+2q)$.

In the weak-coupling phase ($\kappa<\sqrt{\omega_1\omega_2}$), we have the steady state in the local basis
\begin{subequations}
\begin{align}
    &\rho_{11} = \frac{1}{R}(1+p)(1+q);\\
    &\rho_{22} = \frac{1}{R}\left(\sin^2\frac{\theta}{2}p+\cos^2\frac{\theta}{2}q+pq\right);\\
    &\rho_{33} = \frac{1}{R}\left(\cos^2\frac{\theta}{2}p+\sin^2\frac{\theta}{2}q+pq\right);\\
    &\rho_{44} = \frac{pq}{R};\\
    &\rho_{23} = \frac{\sin\theta}{2R}\left(p-q\right).
\end{align}
\end{subequations}

In the strong-coupling phase ($\kappa<\sqrt{\omega_1\omega_2}$), we have the steady state in the local basis
\begin{subequations}
\begin{align}
    &\rho_{11} = \frac{1}{R}(1+p)q;\\
    &\rho_{22} = \frac{1}{R}\left(\cos^2\frac{\theta}{2}\left(1+p+q\right)+pq\right);\\
    &\rho_{33} = \frac{1}{R}\left(\sin^2\frac{\theta}{2}\left(1+p+q\right)+pq\right);\\
    &\rho_{44} = \frac{1}{R}(1+q)p;\\
    &\rho_{23} = -\frac{\sin\theta}{2R}\left(1+p+q\right).
\end{align}
\end{subequations}

In the fermionic equilibrium setup, we have the parameters
\begin{equation}
    p = \frac{1}{e^{\beta(\varepsilon_+-\mu)}+1},\quad q = \frac{1}{e^{\beta(\varepsilon_--\mu)_-}+1}.
\end{equation}
The steady state in the energy basis is diagonal, given by
\begin{subequations}
\begin{align}
    &\rho_{gg} = (1-p)(1-q);\\
    &\rho_{e_1e_1} = (1-p)q;\\
    &\rho_{e_2e_2} = (1-q)p;\\
    &\rho_{e_3e_3} = pq.
\end{align}
\end{subequations}
We only consider the weak-coupling phase in the fermionic setup. The steady state at the local basis has the form
\begin{subequations}
\begin{align}
    &\rho_{11} = (1-p)(1-q);\\
    &\rho_{22} = \sin^2\frac{\theta}{2}p+\cos^2\frac{\theta}{2}q-pq;\\
    &\rho_{33} = \cos^2\frac{\theta}{2}p+\sin^2\frac{\theta}{2}q-pq;\\
    &\rho_{44} = pq;\\
    &\rho_{23} = \frac {\sin\theta} {2}(p-q).
\end{align}
\end{subequations}

%\bibliographystyle{apsrev4-2}
%\bibliography{Steer}

\providecommand{\noopsort}[1]{}\providecommand{\singleletter}[1]{#1}%

\end{document}